\documentclass[%
 reprint,
superscriptaddress,
 amsmath,amssymb,
 aps,
]{revtex4-2}

\usepackage{graphicx}% Include figure files
\usepackage{dcolumn}% Align table columns on decimal point
\usepackage{bm}% bold math
\usepackage{lipsum}
\usepackage{xcolor}

\begin{document}

\title{Dissociation dynamics of a diatomic molecule in an optical cavity}% Force line breaks with \\

\author{Subhadip Mondal}
\affiliation{%
 Department of Chemistry,
 Indian Institute of Technology, Kanpur, Uttar Pradesh 208 016, India
}%

\author{Derek S. Wang}
\affiliation{Harvard John A. Paulson School of Engineering and Applied Sciences, Harvard University, Cambridge, MA 02138, USA}

\author{Srihari Keshavamurthy}%
\email{srihari@iitk.ac.in}
\affiliation{%
 Department of Chemistry,
 Indian Institute of Technology, Kanpur, Uttar Pradesh 208 016, India
}%


\date{\today}% It is always \today, today,

\begin{abstract}
 We study the dissociation dynamics of a diatomic molecule, modeled as a Morse oscillator, coupled to an optical cavity.  A marked suppression of the dissociation probability, both classical and quantum, is observed for cavity frequencies significantly below the fundamental transition frequency of the molecule.  We show that the suppression in the probability is due to the nonlinearity of the dipole function.  The effect can be rationalized entirely in terms of the structures in the classical phase space of the model system.
\end{abstract}

\maketitle

%\tableofcontents

\section{\label{sec:intro} Introduction}

Recent experiments in polariton chemistry \cite{thomas2016ground, thomas2019tilting, lather2019cavity, vergauwe2019modification, thomas2020ground,nagarajan2021chemistry} suggests that the quantum nature of light in the cavity quantum electrodynamics (cQED) regime \cite{haroche2006exploring} may play a crucial role in controlling chemistry. These experiments show modified ground-state chemical reactivity of molecules in cavities in the vibrational strong coupling (VSC) regime by tuning the mode frequency of an optical Fabry-Per\'{o}t cavity. An important goal then is to overcome what is believed to be the bane of mode-specific chemistry --- intramolecular vibrational energy redistribution (IVR) \cite{bloembergen1984energy,nesbitt1996vibrational,uzermillerphysrep}---by bringing the cavity mode frequency into resonance with specific vibrational modes of the reactant molecules.  Indeed, several recent studies in the context of VSC have emphasized the role of IVR. We mention a few examples. Sch\"{a}fer \textit{et al.} have shown\cite{schafer2021shining} that the cavity mode can alter the cavity-free IVR pathways leading to the inhibition of a reaction. Chen \textit{et al.} have argued\cite{chen2022cavity} that exciting polariton modes can lead to an acceleration of IVR, whereas there is little change in the dynamics upon exciting the dark modes. In the collective regime, Wang \textit{et al.} demonstrate\cite{wang2022chemical} that under suitable conditions there can be enhanced vibrational energy flow into the cavity mode and thus resulting in slowing down of unimolecular reactions. The importance of vibrational anharmonicity  has been emphasized\cite{hernandez2019multi} by Hern\'{a}ndez and Herrera in terms of the formation of vibrational polaritons exhibiting a bond strengthening effect.  Although the crucial role of IVR in VSC is being increasingly appreciated, the mechanism by which the cavity modulates the free molecule IVR pathways is not yet clear. Therefore, given that the cavity mode corresponds to a harmonic oscillator, one expects that our current understanding\cite{gruebele2004vibrational,leitner2015quantum,karmakar2020intramolecular,farantos2009energy} of IVR in isolated molecules will be relevant in the context of polariton chemistry as well.

A firm theoretical understanding of VSC, particularly in the experimentally relevant limit of a large number of molecules in the cavity, is still far from established. Nevertheless, several theoretical studies have provided insights into the possible mechanisms by which the reaction rates may get influenced in the VSC regime \cite{campos2019resonant,li2021cavity,li2021theory,schafer2021shining,du2022catalysis,mandal2022theory,wang2022ivr,yang2021quantum}. It is now well understood that the transition state theory (TST) without dynamical corrections is not capable of explaining the experimental observations \cite{galego2019cavity,li2020origin,campos2020polaritonic,zhdanov2020vacuum}. However, a comparison of the TS recrossings in terms of the dynamical correction factor $\kappa$ (transmission coefficient) in the presence ($\kappa_{c}$) and absence ($\kappa_{0}$) of the cavity  indicates $\kappa_{c} < \kappa_{0}$. Thus, reaction rates are typically reduced upon tuning the cavity mode frequency, while some experiments demonstrate rate acceleration. Given the form of the Pauli-Fierz Hamiltonian (see below) and that the cavity mode is a simple harmonic oscillator, conventional multidimensional reaction rate theories can also be brought to bear on the issue \cite{li2021cavity}. Indeed, tools and concepts based on gas phase TST to the condensed phase rate theories of Grote-Hynes and Kramers  have been invoked \cite{li2021cavity,sun2022suppression,lindoy2022resonant,philbin2022chemical}. Studies utilizing models based on the quantized Jaynes-Cummings \cite{ribeiro2018polariton} and Tavis-Cummings model \cite{botzung2020dark,zhoutaviscummings2022,gera2022effects} that treat molecules as harmonic oscillators, classical molecular dynamics simulations \cite{li2020cavitywater,li2021collective,li2022polaritonmd} and rigorous \textit{ab initio} path integral studies \cite{li2022pimd} have also been performed to uncover the potential mechanisms.  Nevertheless, despite the large number of studies, the theoretical results are still inconclusive; we refer the reader to the recent reviews \cite{nagarajan2021chemistry,wang2021roadmap,li2022review,sidler2022perspective,mandal2022theoretical,yuenzhou2022theoretical} for a summary of the progress till now. Note that even experimentally there are concerns about the correct interpretation of the observed effects \cite{imperatore2021reproducibility}.  

A promising approach for further study is quantum dynamics simulations of cavity-molecule systems that fully describe the anharmonic nature of molecular vibrations. While these single-molecule models do not capture the complexity of collective coupling in VSC experiments, similar models have proven invaluable for understanding molecule-light interactions and can shed light on polariton chemistry. For instance, in 1977, Miller introduced the Hamiltonian for a single cavity mode interacting with a diatomic represented by a Morse oscillator to provide a consistent semiclassical description for absorption, induced emission, and spontaneous emission processes \cite{miller1978classical}. The diatomic molecule has a single vibrational degree of freedom and hence issues associated with IVR within the molecule do not arise, allowing one to focus solely on the influence of the molecule-cavity energy flow dynamics on the reactivity. Along these lines, the Morse oscillator-cavity model provided valuable classical and semiclassical insights into the excitation and dissociation in diatomic systems \cite{brown1986quantum,shirts1984cm,davis1982surface}, although these studies do not include the crucial dipole self energy (DSE) term. Recently, Fischer and Saalfrank performed a detailed quantum study of the Morse oscillator-cavity system using the Pauli-Fierz Hamiltonian with the DSE included \cite{fischer2021ground}. They conclude that, despite the formation of vibrational polaritons, there is no substantial change in the dissociation energies and bond lengths for coupling strengths even beyond the VSC threshold. However, dissociation dynamics of the diatomic molecule was not studied.

Therefore, in the present work we study a simple model system consisting of a single diatomic molecule, modeled as a Morse oscillator, coupled to a harmonic cavity mode.  The aim is to explore whether the dissociation dynamics is influenced by coupling to the cavity. Both classical and quantum dynamical studies are done in the VSC regime to assess the relevance of quantum effects. Our results demonstrate significant cavity mode frequency-dependent suppression of the dissociation probability. Notably, maximal suppression occurs when the cavity mode frequency is tuned not to the fundamental transition frequency of the diatomic molecule, but rather to far red-shifted frequencies. Interestingly, both the classical and the quantum dissociation probabilities exhibit this modulation around the same cavity mode frequency range At these red-shifted frequencies, certain key nonlinear resonances in the classical phase space of the molecule disappear. Such resonances, involving the molecular vibration and the cavity mode, are responsible for energy exchange between the molecule and the cavity i.e., molecule-cavity IVR. Our analysis reveals that the nonlinearity of the molecular dipole function plays a crucial role. In fact, within the linearized dipole approximation the dissociation dynamics shows very little modulation over a wide range of the cavity frequencies. 

The paper is organized as follows. In Sec.~\ref{sec:model} we give details on the model Hamiltonian along with the relevant parameters utilized in this work. The classical and quantum dissociation probabilities in the VSC regime are compared in Sec.~\ref{sec:dissprobs}, illustrating the essential role of the dipole function. A classical phase space based understanding of the results is presented in Sec.~\ref{sec:phasespace}, followed by an analysis of the results in Sec.~\ref{sec:analysis} and conclusions and future directions in Sec.~\ref{sec:disc}.

\section{\label{sec:model} Model Hamiltonian}
Our model system corresponds to a diatomic molecule coupled to a quantized electromagnetic field mode of a Fabry-P\'{e}rot cavity. According to Fischer and Saalfrank \cite{fischer2021ground}, the Pauli-Fierz Hamiltonian within the dipole approximation and in the length gauge can be expressed as $H = H_{M} + H_{C} + H_{MC}$ 
\begin{subequations}
\begin{align}
H_{M}(q,p) &= \frac{1}{2m}p^{2} + D\left(1-e^{-\alpha(q-q_{e})}\right)^2 \\
H_{C}(q_{c},p_{c}) &= \frac{1}{2} \left(p_{c}^{2} + \omega_{c}^{2} q_{c}^{2} \right) \\
H_{MC}(q,q_{c}) &= \omega_{c} \lambda_{c} q_{c} \mu(q) + \frac{1}{2} \lambda^{2}_{c}\mu^{2}(q)  
\end{align}
\end{subequations}

\begin{center}
\begin{table}[t]

    \caption{Parameters for the HF molecule \cite{brown1986barriers}}
    \vspace{0.1in}
\centering
\begin{tabular}{|c||l|l|}
\hline
 Symbol &  Description &  Value (in au) \\
\hline
$\alpha$ & Morse parameter of HF Bond & 1.174  \\
$D$ & Dissociation energy & 0.225 \\
$q_e$ & Equilibrium  bond length & 1.7329 \\
$m$ & Reduced mass   & 1744.59  \\
$A$ & Dipole moment parameter & 0.4541 \\
$B$ & Dipole moment parameter & 0.0064  \\
$\mu^{\prime}$ & Dipole moment derivative &  0.33  \\
\hline
\end{tabular}\label{tbl:HF_parameter}
\end{table}
\end{center}
where ${\bf Q} = (q,q_{c})$ and ${\bf P} = (p,p_{c})$ are the dynamical position and conjugate momentum variables. We have assumed a single cavity mode, described by the harmonic Hamiltonian $H_{C}$ with frequency $\omega_{c}$, that is polarized along the molecular axis. In addition, we take an ideal cavity with no loss. The molecular Hamiltonian  is denoted by $H_{M}$ and the vibration of the diatomic molecule is modeled by a Morse oscillator with $q_{e}$ and $D$ denoting the equilibrium bond length and the dissociation energy respectively.  The first term of $H_{MC}$ is the molecule-cavity coupling ($H_{int}$), characterized by a coupling strength $\lambda_{c} \equiv (\epsilon_{0} \epsilon_{r} {\cal V})^{-1/2}$ with $\epsilon_{0},\epsilon_{r}$, and ${\cal V}$ being the vacuum dielectric constant, dielectric permittivity,and cavity volume respectively. In this work, we set $\epsilon_{r}=1$. The second term in $H_{MC}$ is the  dipole self energy (DSE). We include the DSE in all our computations, classical and quantum, since several studies have established its importance for a proper analysis of the coupled cavity-molecule dynamics \cite{fischer2021ground,schafer2020relevance}. Note that the Hamiltonian $H({\bf Q},{\bf P})$  is  a two degrees-of-freedom autonomous system that conserves the total cavity-molecule energy.

\begin{figure*}[htbp]
    \centering
    \includegraphics[width=0.95\textwidth]{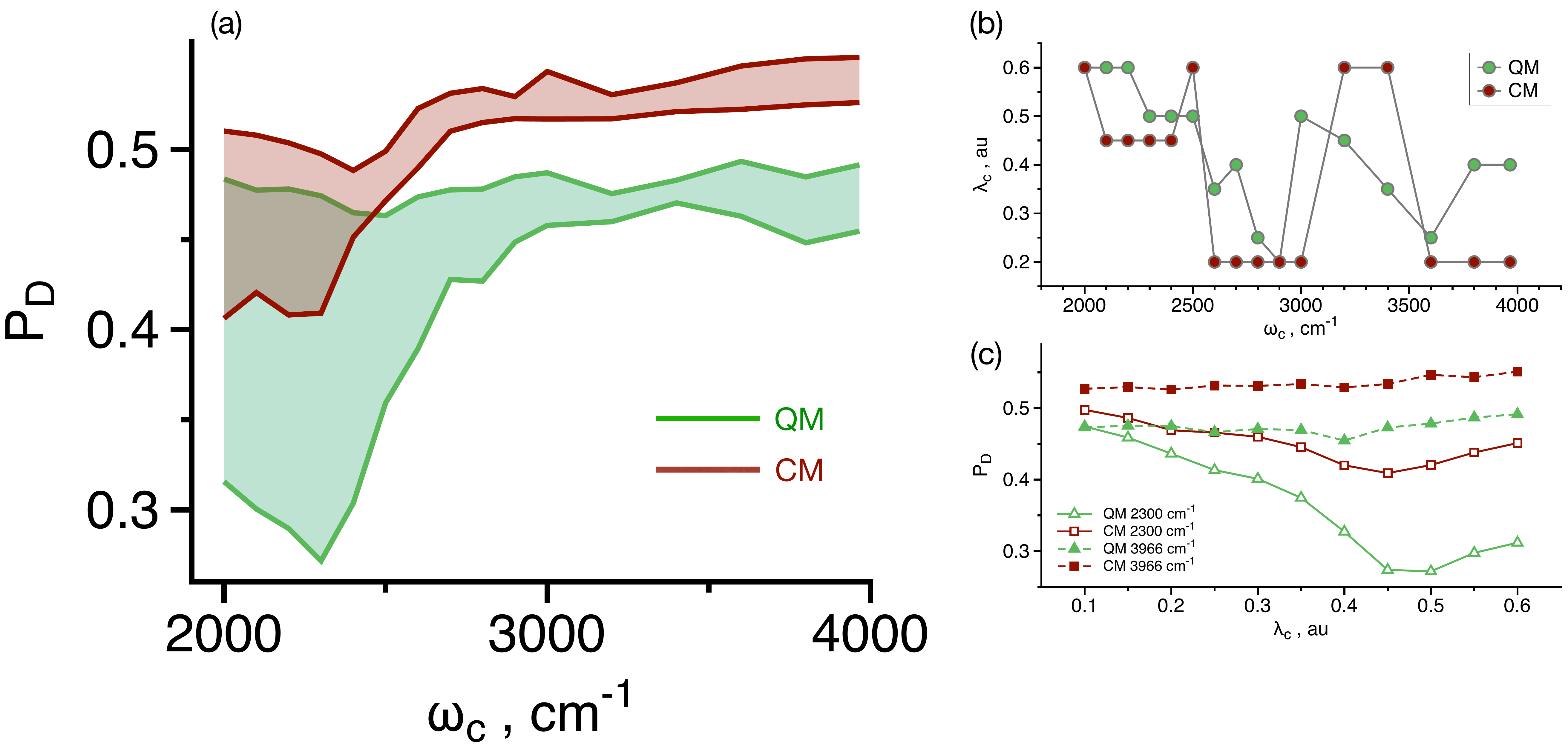}
    \caption{(a) Classical and quantum dissociation probabilities $P_{D}$ as a function of the cavity frequency $\omega_{c}$ and the molecule-cavity coupling strength $\lambda_{c}$. The shaded regions indicate the variation of the dissociation probabilities for $\lambda_{c} \in (0.1,0.6)$ au incremented in steps of $0.05$. (b) The values $(\omega_{c},\lambda_{c})$ for which maximum suppression in $P_{D}$ occurs. (c) Classical and quantum  $P_{D}$ versus $\lambda_{c}$ for two example cavity frequencies $\omega_{c} = 2300$  cm$^{-1}$ and $\omega_{c} = \omega_{01} \approx 3966$ cm$^{-1}$.}
    \label{fig:dissocprob_params}
\end{figure*}

Following the earlier work we also define the parameters 
\begin{equation}
g = \sqrt{\frac{\hbar\omega_{c}}{2}}\lambda_{c} \label{g_lambdac} \,\,\,\,\, ; \,\,\,\,\, \Omega_{R} = \frac{2g \sqrt{N} |d_{fi}|}{\hbar}
\end{equation}
with $g$ (V/m in SI units) being a measure of the cavity-molecule interaction strength. The Rabi frequency $\Omega_{R}$ for $N$ molecules ($N = 1$ in this work) in the cavity is expressed in terms of the transition dipole moment $d_{fi}$ between the initial ($i$) and final ($f$) vibrational states of the molecule. 
The dimensionless parameter $\eta \equiv \Omega_{R}/2\omega_{c} \approx \mu^{\prime}(q_{e}) \lambda_{c} (4m \omega_{0} \omega_{c})^{-1/2}$ (within the linearized dipole approximation) determines the specific coupling regime we are in; by convention $\eta \ge 0.1$ marks the transition from the vibrational strong coupling (VSC) to the vibrational ultrastrong coupling (VUSC) regime. 
In this work we use the following functional form \cite{stine1979classical} of the dipole moment function  
\begin{equation}
   \mu(q)=A q e^{-B q^4} \label{dipol_nonlin}
\end{equation}
For future reference we provide the linear approximation to the dipole function: 
\begin{equation}
    \mu(q)=\mu(q_{e})+\left [\frac{d\mu(q)}{dq} \right]_{q_{e}}(q - q_{e}) \label{dipol_lin}
\end{equation}
The parameters for the diatomic molecule, taken from an earlier work \cite{brown1986barriers} by Brown and Wyatt, are chosen to represent the hydrogen fluoride (HF) molecule and  are given in Table~\ref{tbl:HF_parameter}. 

\begin{figure*}[htbp]
    \centering
    \includegraphics[width=0.95\textwidth]{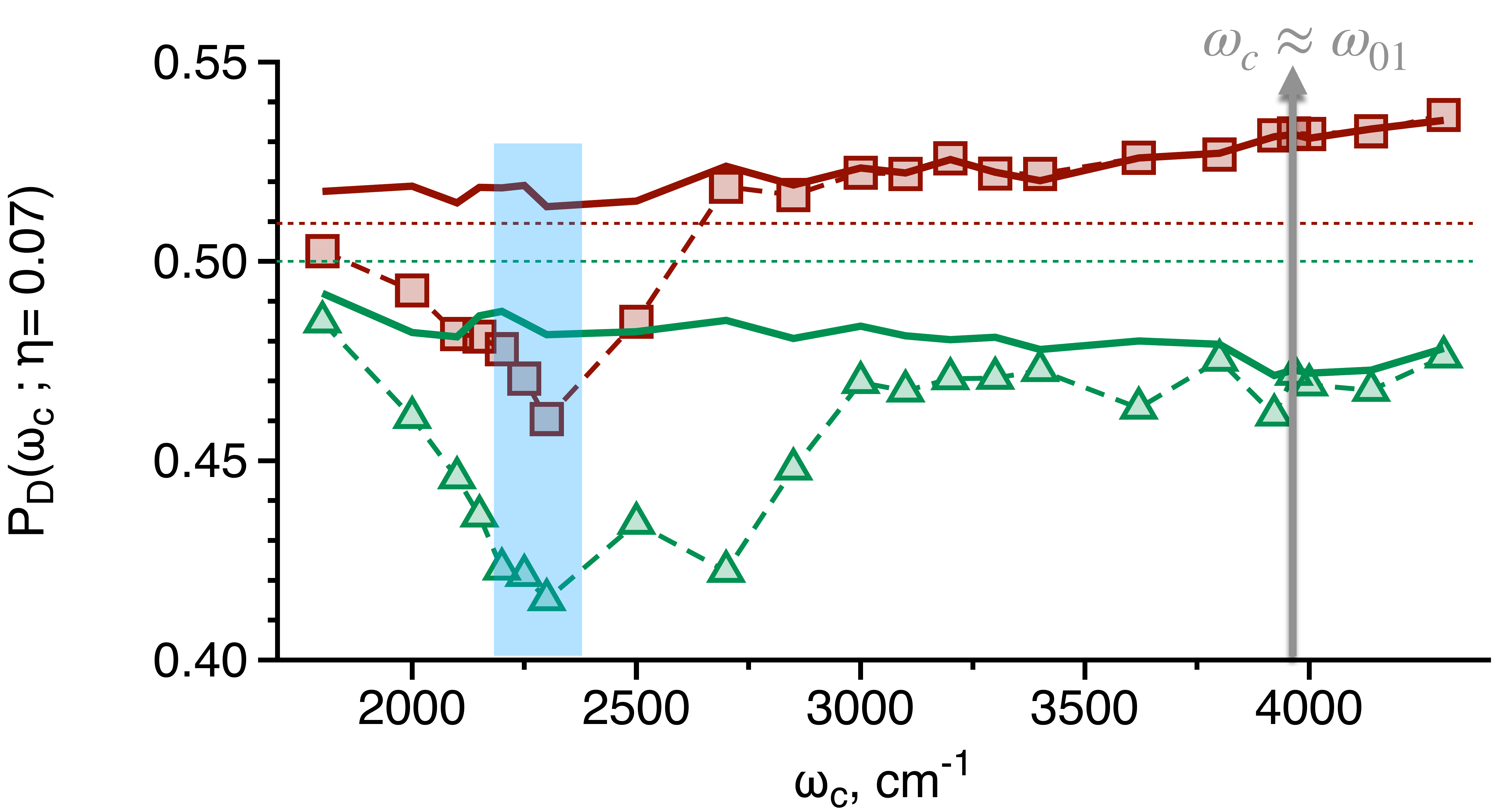}
    \caption{Dissociation probability of HF as a function of the cavity frequency. Total energy of the system (cavity $+$ molecule) is $E = 0.25$ au and the parameter $\eta = 0.07$ is fixed. The quantum (green triangles) and the corresponding classical (brown squares) results, utilizing the dipole function in eq.~\ref{dipol_nonlin}, are shown. For comparison, the results are also shown for the linear dipole approximation (thick lines) and cavity - free limit (dashed lines). The grey vertical line indicates the cavity frequency being resonant with the HF $0 \rightarrow 1$ fundamental. Note the blue shaded region exhibiting significant suppression of the dissociation probability, with very good classical-quantum correspondence. See text for details and discussions. }
    \label{fig:dissocprob_master}
\end{figure*}

\section{\label{sec:results} Results and discussion}

\subsection{\label{sec:dissprobs} Dissociation probability: classical and quantum}

In order to study the dissociation dynamics of the diatomic molecule, the initial state is chosen to be a  polariton wavepacket \cite{fischer2021ground}
\begin{equation}
    \Psi_{0}(q,q_{c};t=0) = \psi_{G}(q;q_{0}) \phi_{0}(q_{c}) \label{initial_wavefunc}
\end{equation}
with 
\begin{equation}
    \psi_{G}(q;q_{0}) = \left(\frac{1}{\pi\sigma^{2}}\right)^{1/4}\exp\left[-\frac{(q-q_{0})^{2}}{2\sigma^{2}}\right] \label{HF_Gaussian}
\end{equation}
being a displaced ground state wavefunction of the harmonized HF bond of frequency $\omega_{0} \approx 4139$ cm$^{-1}$ and width $\sigma \equiv (\hbar/m \omega_{0})^{1/2}$. We consider the cavity to have  no photons initially and hence the state $\phi_{0}(q_{c})$ is taken to be the ground harmonic eigenstate of the cavity mode. The center of the wavepacket $q_{0}$  is then chosen such that $\langle H \rangle_{\Psi_{0}} \equiv \langle \Psi_{0}| H | \Psi_{0} \rangle$ corresponds to the desired total energy. Here we fix $\langle H \rangle_{\Psi_{0}} = 0.25$ au which is above the dissociation energy of the HF bond. Note that qualitatively similar results are obtained for other values of $\langle H \rangle_{\Psi_{0}}$ as well (see Fig.~\ref{fig:dissocprob_master} and Fig.~\ref{fig:dissprob_moreexamps} below).

The time evolved quantum state $\Psi(q,q_{c};t)$ is obtained by numerically solving (see Appendix) the Schr\"{o}dinger equation and the quantum dissociation probability is then calculated as
\begin{equation}
    P^{\rm QM}_{D}(t) = 1 - \langle \Psi(q,q_{c},t) | \Psi(q,q_{c},t) \rangle \label{QM_pd}
\end{equation}
The corresponding classical dissociation probabilities $P^{\rm CM}_{D}(t)$ are computed (see Appendix) by choosing an ensemble of initial conditions $N_{\rm tot}$ sampled from the classical density $\rho_{cl}(q,p,q_{c},p_{c},0)$ corresponding to the initial quantum polaritonic wavepacket. For this study, we take $N_{\rm tot} = 50000$  and  time evolve  each initial phase space point by integrating the Hamiltonian's equations of motion. A trajectory is considered to be dissociated when the displacement $(q-q_{e}) \geq 7.5$ au.  The classical dissociation probability is calculated as 
\begin{equation}
    P^{\rm CM}_{D}(t) = \frac{N_{\rm diss}(t)}{N_{\rm tot}} \label{CM_pd}
\end{equation}
with $N_{\rm diss}$ being the number of dissociated trajectories.

In Fig.~\ref{fig:dissocprob_params} the results of our computations are shown for a range of cavity frequencies $\omega_{c}$ and molecule-cavity coupling strengths $\lambda_{c}$ values. The total energy of the molecule-cavity system is fixed at $E = 0.25$ au, which is higher than the dissociation energy of the diatomic molecule. Several observations can be made at this stage. Firstly,  as evident from Fig.~\ref{fig:dissocprob_params}(a), $P^{\rm QM}_{D}$ and $P^{\rm CM}_{D}$ variations are qualitatively similar over the entire range with the classical probability being consistently higher than the quantum case. Secondly, Fig.~\ref{fig:dissocprob_params}(a) reveals  a significant dip in the dissociation probabilities around $\omega_{c} \approx 2200-2400$ cm$^{-1}$, which lies far from the HF fundamental ($\omega_{01}\approx 3966$ cm$^{-1}$) frequency. Note that, for a fixed $\omega_{c}$, both classical and quantum dissociation probabilities exhibit considerable variations with changing  $\lambda_{c}$ in the low cavity frequency regimes. It is also interesting to note from Fig.~\ref{fig:dissocprob_params}(b) that the values of $\lambda_{c}$ which result in the maximum suppression of the dissociation probability for a given $\omega_{c}$ exhibits a non-monotonic behavior. In Fig.~\ref{fig:dissocprob_params}(c) the variations of $P_{D}$ with $\lambda_{c}$  are shown for two specific cavity frequencies $\omega_{c} = 2300$ cm$^{-1}$ and $3966$ cm$^{-1}$, with the latter corresponding to the HF fundamental transition. Clearly, the dissociation probability for $\omega_{c}=3966$ cm$^{-1}$ is nearly constant over the entire range of the molecule-cavity coupling values.

In order to bring out further details, Fig.~\ref{fig:dissocprob_master} shows the quantum and the classical dissociation probabilities as a function of $\omega_{c}$ for a fixed value of the dimensionless coupling $\eta = 0.07$, corresponding to typical VSC regimes, and $E = 0.25$ au. Furthermore, note that $\omega_{c} = \omega_{01} \approx 3966$ cm$^{-1}$ corresponds to $\lambda_{c} \approx 0.33$ au ($\eta = 0.07$) and a Rabi splitting $\Omega_{R} \approx 530$ cm$^{-1}$. For reference, we also show the uncoupled (outside the cavity) results. The specific parameters considered here are representative and similar results are obtained for slightly different choice of the $(E,\eta)$ parameters. As an example, Fig.~\ref{fig:dissprob_moreexamps} shows the variation in $P_{D}$ with $\omega_{c}$ for different initial total energy values.
For both the classical and quantum cases one can clearly observe a significant difference between the probabilities when utilizing the non-linear dipole function in eq.~\ref{dipol_nonlin} versus the linearized dipole approximation. In particular, the dip in the dissociation probabilities around $\omega_{c} \approx 2200-2400$ cm$^{-1}$ is absent  within the linear dipole approximation. It is also useful to note that $P^{\rm QM}_{D}$ is below the uncoupled value over the entire range of cavity frequencies considered, whereas for $\omega_{c} \geq 2750$ cm$^{-1}$ the classical results lie slightly above the uncoupled limit. Finally, consistent with the remarks made above and irrespective of using the actual dipole function or the linearized approximation, there is hardly any change in the dissociation probabilities when the cavity frequency is tuned to the fundamental vibrational transition of the diatomic molecule. 

\begin{figure}[htbp]
    \centering
    \includegraphics[width=0.50\textwidth]{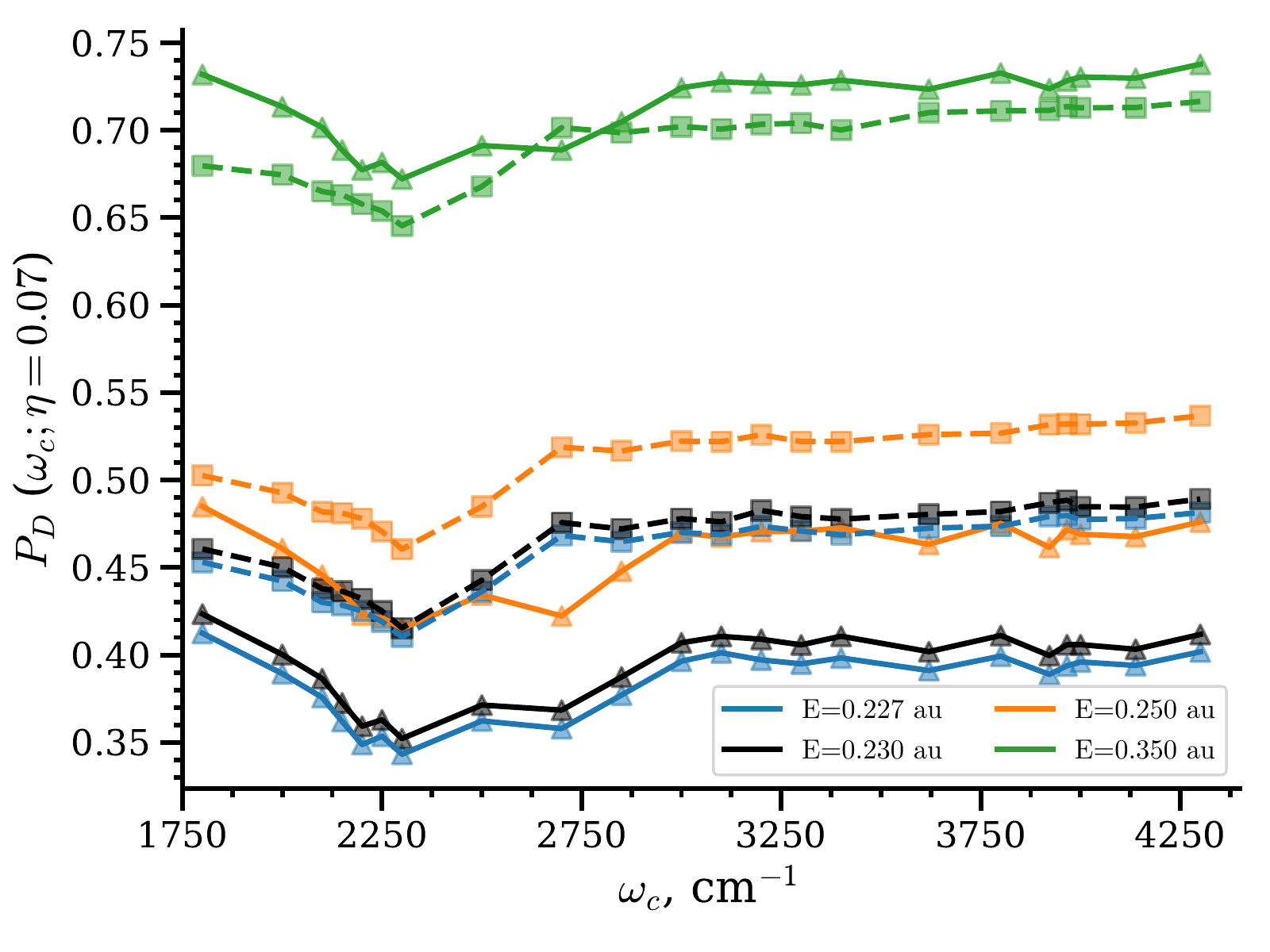}
    \caption{Dependence of the dissociation probability on the cavity frequency for various initial total energy values. The coupling strength is fixed at $\eta = 0.07$ and both classical (squares) and quantum (triangles) results are shown.}
    \label{fig:dissprob_moreexamps}
\end{figure}

\begin{figure*}[htbp]
    \centering
    \includegraphics[width=0.95\textwidth]{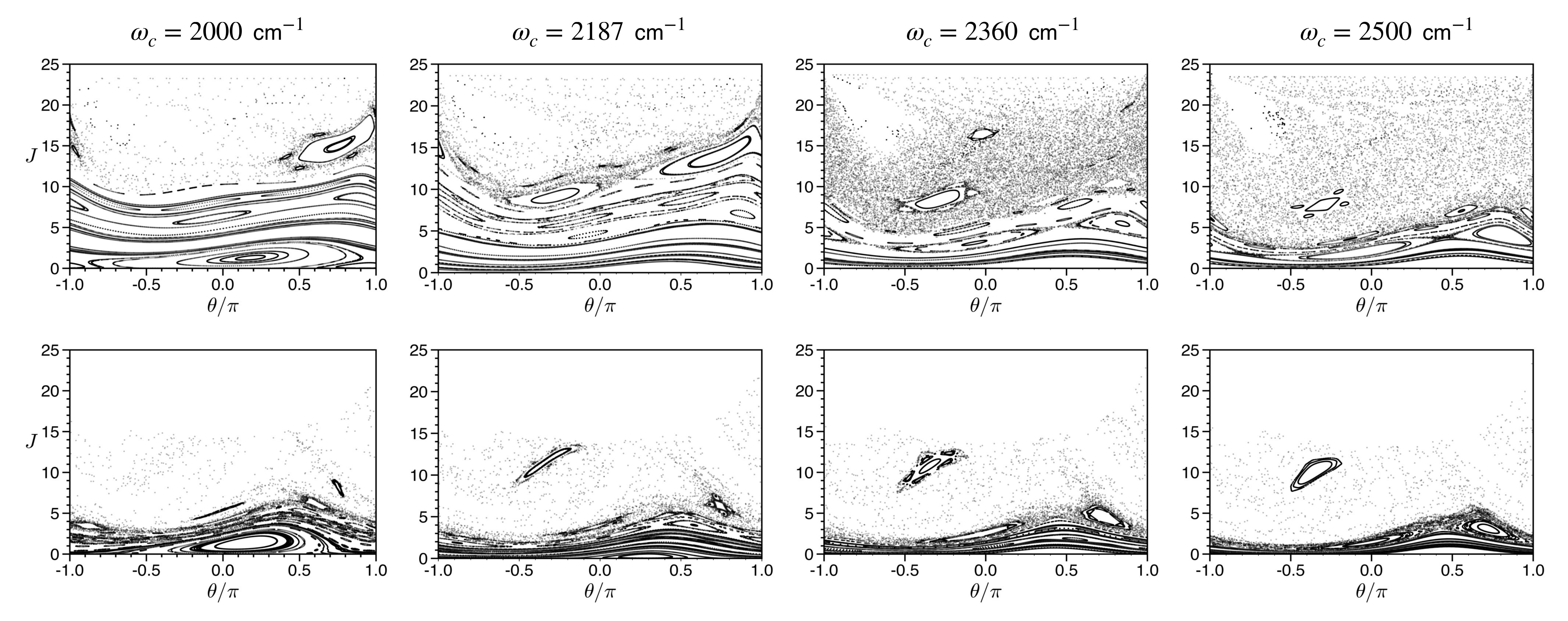}
    \caption{Phase space $(J,\theta)$ surface of section of the molecule-cavity system with varying cavity frequency, utilizing the non-linear dipole function (top row, eq.~\ref{dipol_nonlin}) and a linear approximation (bottom row, eq.~\ref{dipol_lin}). The total energy $E = 0.25$ au and the light-matter coupling strength $\eta = 0.07$ are fixed. The cavity mode frequency $\omega_{c}$ are indicated.}
    \label{fig:j_th_eta0.07}
\end{figure*}

\subsection{\label{sec:phasespace} Classical phase space: importance of the nonlinearity of the dipole}

Given that $P^{\rm CM}_{D}$ in Fig.~\ref{fig:dissocprob_master} also exhibits a dip we surmise that the mechanism is of classical origin. Indeed, early semiclassical studies by Lima and coworkers \cite{lima2013cm,lima2014qm} on the driven Morse oscillator has established that significant variations in the dissociation probabilities can arise from the nonlinearity of the dipole function, and more recently, Triana \textit{et al.} argue that the shape of the dipole function is crucial for understanding the dynamical properties of vibrational polaritons \cite{triana2020shape}. In what follows we illustrate the effect for the cavity-Morse model of interest.

Thus, we analyse the detailed classical dynamics to inspect changes in the phase space structures with varying cavity mode frequency. In the context of driven Morse oscillator dynamics it is well known that the classical phase space structures such as chaos and the various field-matter nonlinear resonances regulate the dissociation dynamics. In the present study, the system as described by the Pauli-Fierz Hamiltonian is a two dimensional conservative system and hence the dynamics can be conveniently monitored by computing Poincar\'{e} surface of section (PSOS) at fixed total energy $E$ and varying $\omega_{c}$.  Presently, we take $q_{c}=0$, $p_{c} > 0$ and  $q=0$, $p > 0$ as sectioning planes for the computation of ($q,p$) and ($q_{c},p_{c}$) PSOS respectively. 

The results of our PSOS computations are shown in Fig.~\ref{fig:j_th_eta0.07} over the relevant range of cavity frequencies. Note that the phase spaces are shown in terms of the known action-angle variables $(J,\theta)$ for the Morse oscillator and $(J_{c},\theta_{c})$ for the harmonic oscillator corresponding to the cavity mode. Since semiclassically the actions correspond to quantum numbers, the PSOS in such variables yield direct information on the extent of excitation in the two modes. Note that the top panel of Fig.~\ref{fig:j_th_eta0.07} uses the full nonlinear dipole function whereas the bottom panel of Fig.~\ref{fig:j_th_eta0.07} utilizes the linearized dipole function. On inspecting the phase spaces it is immediately clear that there are significant differences between the actual dipole and the linearized cases.  Clearly, certain  phase space structures seem to disappear and reappear in the nonlinear dipole case at specific cavity frequencies. Moreover, the phase space is less chaotic in the nonlinear dipole case when compared to the linearized case. Interestingly, for $\omega_{c} = 2187$ cm$^{-1}$ one can observe a marked increase in the regularity of the phase space. We also note that the dip in  $P^{\rm CM}_{D}$ seen in Fig.~\ref{fig:dissocprob_master} appears to correlate with the extent of regularity in the phase space.

\begin{figure*}[htbp]
    \centering
    \includegraphics[width=0.95\textwidth]{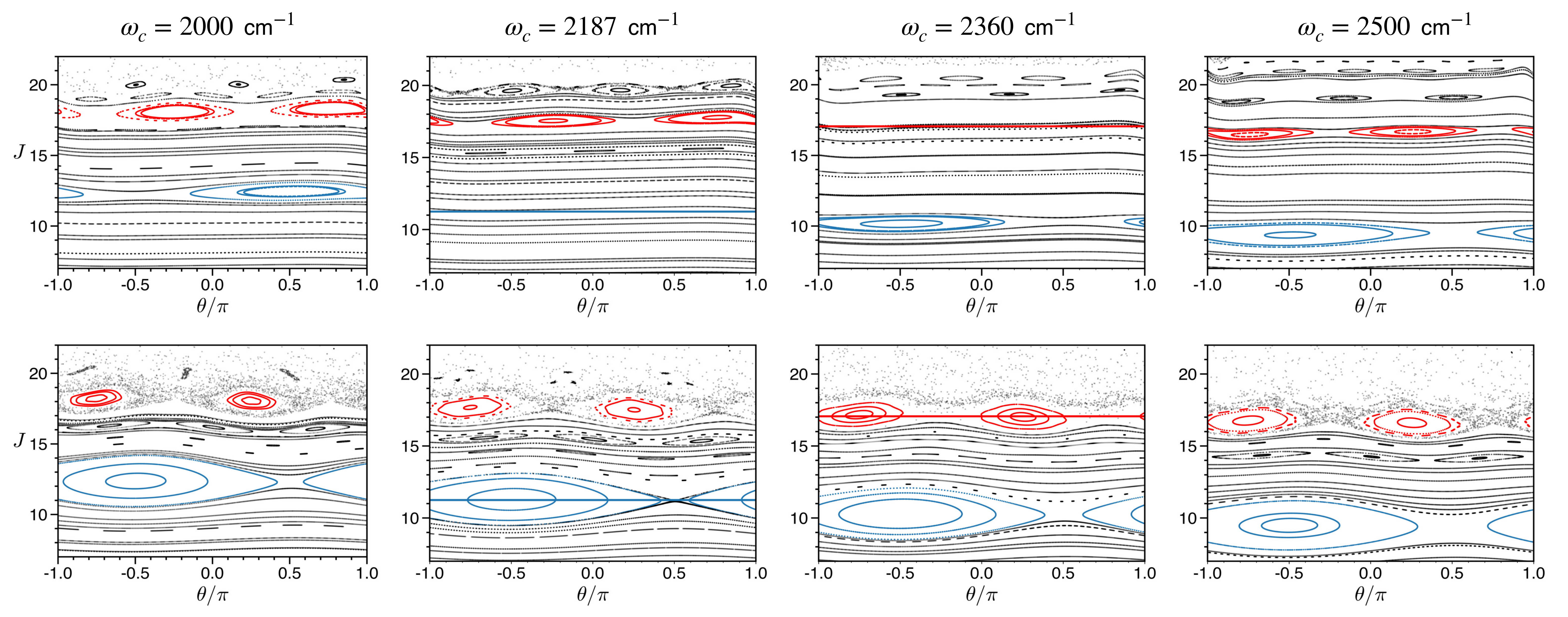}
    \caption{Same as in Fig.~\ref{fig:j_th_eta0.07}, but for a much lower light-matter coupling strength $\lambda_{c}=0.01$ au.  The $1$:$1$ and $2$:$1$ matter-cavity nonlinear resonances are shown in blue and red colors respectively. The solid blue and red lines  for $\omega_{c} = 2187$ and $2360$ cm$^{-1}$ respectively indicate the prediction of eq.~\ref{jn_res} for the location of the $1$:$1$ and $2$:$1$ resonances. Note the near disappearance of the resonances in the top row corresponding to the nonlinear dipole case.}
    \label{fig:j_th_ld_nld}
\end{figure*}

To make the above discussion more explicit, consider the molecule-cavity Hamiltonian $H({\bf J},\bm \theta) = H_{0}({\bf J})+ V({\bf J},\bm \theta)$ in terms of the action-angle variables with
\begin{eqnarray}
    H_{0}({\bf J}) &=& \omega_{0} J \left(1 - \frac{\omega_{0}}{4D} J\right) + \omega_{c} J_{c} \\
    &\equiv& H_{0}^{(M)}(J) + H_{0}^{(C)}(J_{c}) \nonumber \label{Hamiltonian_jjc}
\end{eqnarray}
being the zeroth-order uncoupled Hamiltonian and $\omega_{0} \equiv (2 \alpha^{2}D/m)^{1/2}$ is the harmonic frequency associated with the Morse oscillator. The coupling term can be written as
\begin{eqnarray}
       V({\bf J},\bm \theta) &=& \epsilon \sum_{n=0}^{\infty}\tilde{V}_{n}({\bf J}) \left[\sin(n \theta + \theta_{c}) - \sin(n \theta - \theta_{c}) \right] \nonumber \\
    &+& \frac{1}{2} \lambda_{c}^{2} \left[\sum_{n=0}^{\infty} V_{n}(J)\cos(n \theta) \right]^{2} 
     \label{Hamiltonian_jtheta}
\end{eqnarray}
where we have denoted $\tilde{V}_{n}({\bf J}) \equiv J_{c}^{1/2} V_{n}(J)$ and $\epsilon = \lambda_{c}(\omega_{c}/2)^{1/2}$ is the cavity-molecule coupling. The last term in the above expression comes from the DSE and note that it does not depend on $\epsilon$. However, note that the surface of sections shown in Fig.~\ref{fig:j_th_ld_nld} does include the DSE term.

In Eq.~\ref{Hamiltonian_jtheta}, the term $\sin(n \theta - \theta_{c})$ gives rise to a $n$:$1$ nonlinear resonance in the classical phase space. Specifically, this condition can be expressed at the zeroth-order level as
\begin{equation}
    n \Omega_{0}(J) \equiv n \frac{\partial H_{0}^{(M)}(J)}{\partial J}  = \omega_{c} \equiv \frac{\partial H_{0}^{(C)}(J_{c})}{\partial J_{c}}  \label{nonlinear_res}
\end{equation}
The above $(n+1)^{\rm th}$ order resonance condition is satisfied for the specific resonant oscillator action $J = J_{r}^{(n)}$ 
\begin{equation}
    J_{r}^{(n)} = \frac{2D}{\omega_{0}} \left(1 - \frac{\omega_{c}}{n \omega_{0}}\right) \label{jn_res}
\end{equation}
From the above it is clear that $J_{r}^{(n)}$, the center of a specific order resonance, shifts towards lower action values for increasing $\omega_{c}$, and that the higher order resonances appear at increasingly larger action values for a given $\omega_{c}$. Both of these features can be readily seen in Fig.~\ref{fig:j_th_ld_nld} showing the phase space sections.
The width of a nonlinear resonance of a given order is determined \cite{shirts1984cm} from the Fourier coefficients
\begin{eqnarray}
    V_{n}(J) &=& \frac{1}{\pi} \int_{0}^{\pi}\mu(J,\theta)d\theta  \label{Fourier_0} \,\, ; \, \, n = 0\\
    &=& \frac{2}{\pi} \int_{0}^{\pi}\mu(J,\theta)\cos(n\theta)d\theta \,\, ; \,\, n \neq 0
\end{eqnarray}
In particular, as follows from Chirikov's theory \cite{chirikov1979universal}, the width of a specific order resonance zone scales as $(\epsilon|V_{n}(J_{r}^{(n)})|)^{1/2}$ with higher order resonances having smaller widths. It is well known that the widths play a crucial role in the classical phase space theory of dissociation.  Essentially, overlap of several of the resonances results in large scale chaos and the Morse oscillator dissociates via diffusion through the chaotic phase space. Several studies have also established the close classical-quantum correspondence for this dissociation mechanism and have brought out the role of quantum effects such as localization due to cantori barriers\cite{brown1986barriers,brown1986quantum} and resonance-assisted tunneling\cite{sethi2009local}.  

\begin{figure*}[htbp]
    \centering
    \includegraphics[width=0.95\textwidth]{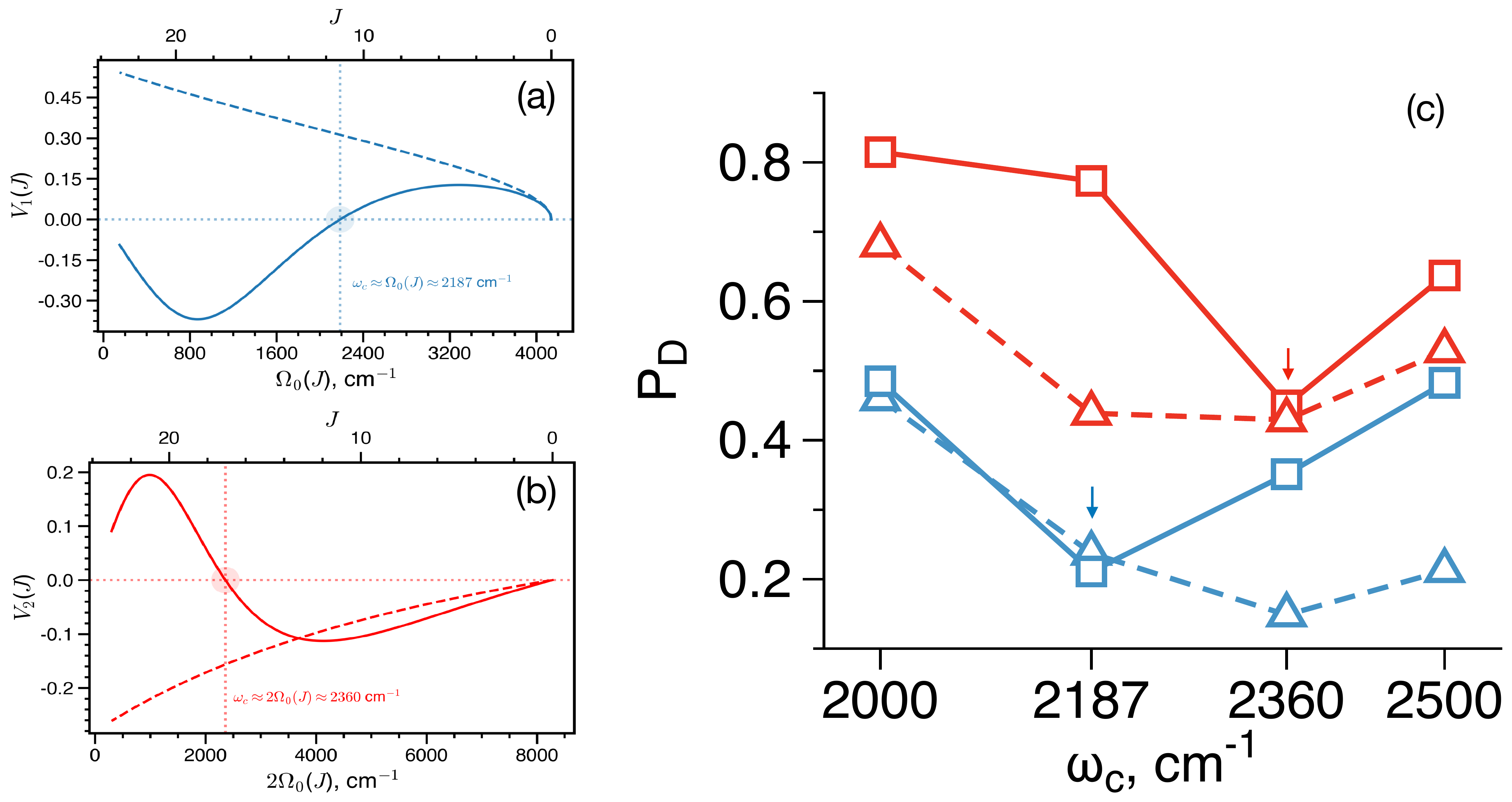}
    \caption{Fourier coefficients, $V_{n}(J)$ versus frequency of the oscillator $n\Omega_{0}(J)$ for (a) $n=1$ (blue) and (b) $n=2$ (red). The dotted vertical line indicates the cavity frequency, $\omega_{c}$ corresponding to the vanishing of $V_{n}(J)$. For comparison, $V_{n}(J)$ values are also shown for the linear dipole approximation (dashed line). (c) Classical (squares) and quantum (triangles) dissociation probabilities for initial Morse oscillator states $v = 11$ (blue) and $v=17$ (red). Parameters $(E,\eta) = (0.25 \text{ au},0.07)$ are fixed and the initial cavity state is determined by the total energy. The arrows indicate the cavity frequency at which the Fourier coefficients vanish.}
    \label{fig:Fourier_coeff_12}
\end{figure*}

\subsection{\label{sec:analysis} Analysis and importance of the cavity-molecule resonances}

Based on the above discussion one surmises that vanishing of a specific Fourier coefficient $V_{n}(J)$ implies that the specific nonlinear resonance has zero width. Hence, that specific pathway for cavity-molecule IVR is no longer available. For the Morse oscillator case the Fourier coefficients for the linear dipole approximation are known \cite{shirts1987use} analytically and it is easy to show that they all vanish only for $\Omega_{0}(J) = \omega_{0}$ i.e., for the ground state. However, as shown by Lima \textit{et al.} \cite{lima2013cm}, for the nonlinear dipole case various Fourier coefficients can vanish at different values of the action. This is confirmed in Fig.~\ref{fig:Fourier_coeff_12}(a) and (b) where we show the vanishing of $V_{1}(J)$ and $V_{2}(J)$ for $\Omega_{0}(J) \approx 2187$ and $2360$ cm$^{-1}$ for the HF parameters. The corresponding resonant action values are $J_{1}^{r} \approx 11.5$ and $J_{2}^{r} \approx 17$, as also indicated in Fig.~\ref{fig:Fourier_coeff_12}. Thus, the $\Omega_{0}(J):\omega_{c} = 1$:$1$ and the $2$:$1$ nonlinear resonances have nearly zero widths for cavity frequencies $\omega_{c}=2187$ and $2360$ cm$^{-1}$ respectively. This switching off of the resonances for specific cavity frequencies can be clearly visualized by inspecting the phase space for smaller coupling strengths. An example is shown in Fig.~\ref{fig:j_th_ld_nld} which clearly shows the disappearance of the $1$:$1$ and $2$:$1$ resonances for $\omega_{c} = 2187$ and $2360$ cm$^{-1}$ respectively. In contrast, note that the linear dipole phase space continues to exhibit the two resonances at $J_{1}^{r} \approx 11.5$ and $J_{2}^{r} \approx 17$.

For coupling strengths in the VSC regime, the vanishing of the Fourier coefficients results in a  more regular phase space in the nonlinear dipole case (cf. Fig.~\ref{fig:j_th_eta0.07}) when compared to the linear dipole phase space. Note that these observations correlate well with the observed dips in the dissociation probability shown in Fig.~\ref{fig:dissocprob_params} and Fig.~\ref{fig:dissocprob_master}. The reason for this is that the initial polariton wavepacket has weights on different vibrational states of the Morse oscillator. Amongst these various states, given the correspondence $J \leftrightarrow (v + 1/2)\hbar$, the classical phase space analysis above predicts that the $v = 11$ and the $v = 17$ states will undergo reduced dissociation. This expectation is confirmed in Fig.~\ref{fig:Fourier_coeff_12}(c) for the energy and coupling strength pertaining to Fig.~\ref{fig:dissocprob_master}. It is clear that dissociation probabilities of these two initial states are reduced by nearly a factor of two for $\omega_{c} = 2187$ and $2360$ cm$^{-1}$. Note that the quantum dissociation probabilities are typically suppressed more compared to the classical values. This is again expected since the increased regularity in the phase space gives rise to sticky regions and partial barriers which tend to localize \cite{brown1986barriers,brown1986quantum} the quantum dynamics to a greater extent. Interestingly, the results in Fig.~\ref{fig:Fourier_coeff_12}(c) show that dissociation can be suppressed even by a ``hot" cavity, observed in Ref.~\citenum{wang2022ivr}.

\begin{figure*}[htbp]
    \centering
    \includegraphics[width=0.95\textwidth]{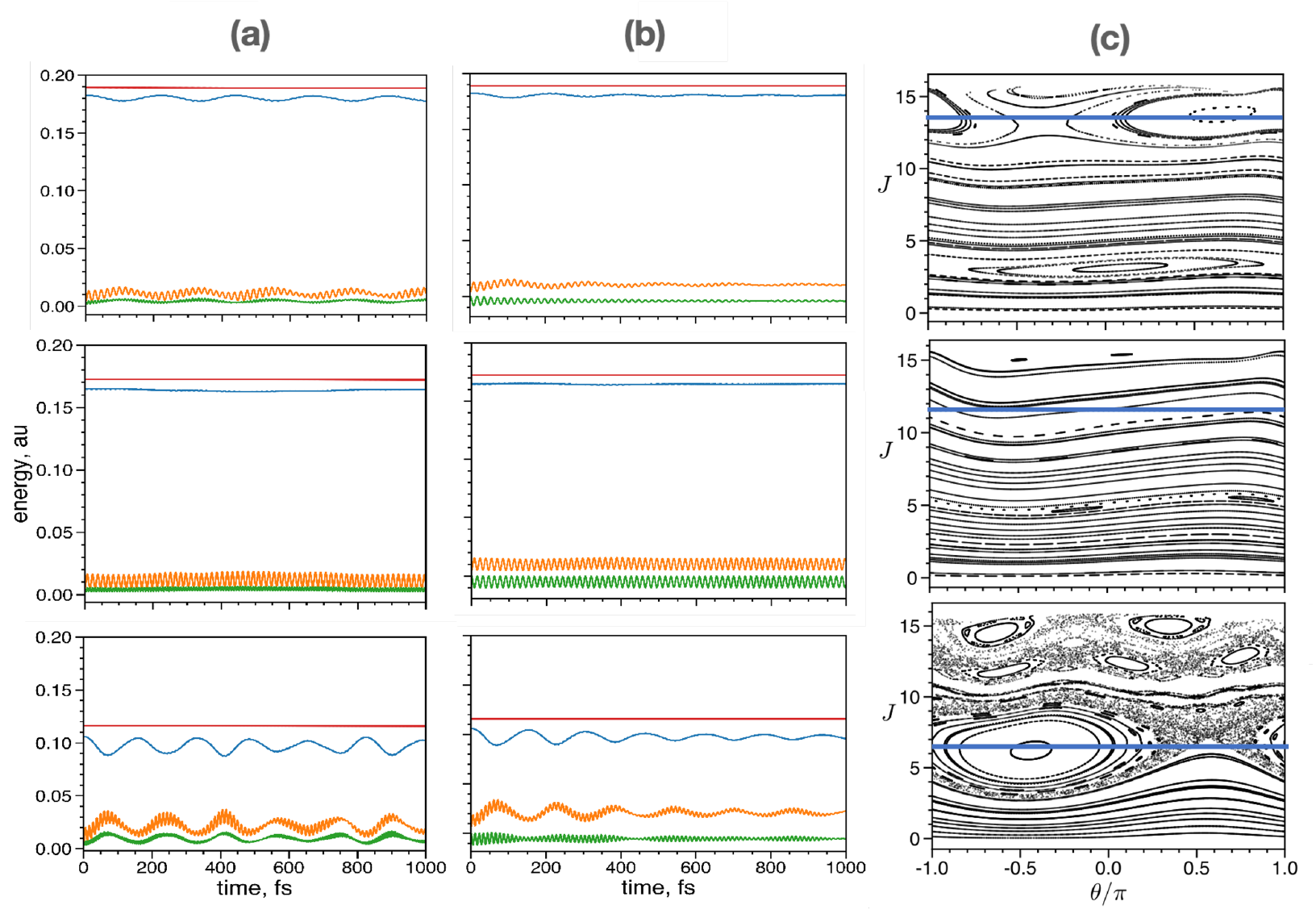}
    \caption{Time dependence of the quantum (a) and the classical (b) energy expectation values of three different initial states for fixed $\lambda_{c} = 0.1$ au. The initial states are $|13,0\rangle$ (top row, $\omega_{c} = 1800$ cm$^{-1}$), $|11,0\rangle$ (middle row, $\omega_{c} = 2187$ cm$^{-1}$), and $|6,0\rangle$ (bottom row, $\omega_{c} = 3000$ cm$^{-1}$). The various expectation values are $\langle H_{M}\rangle$ (blue), $\langle H_{C}\rangle$ (orange), $\langle H_{\rm int}\rangle$ (green), and $\langle H\rangle$ (red). Note that the total energies are different in each case and are below the dissociation threshold. In (c) the corresponding classical phase spaces are shown. The solid blue line indicates the location of the initial Morse action. }
    \label{fig:energy_13_11_6}
\end{figure*}

As a final example emphasizing the role of the vanishing Fourier coefficients in the cavity-molecule IVR pathways, Fig.~\ref{fig:energy_13_11_6} shows the quantum and the classical energy expectation values for specific initial states of the Morse oscillator. The initial actions (vibrational quantum numbers) are chosen such that they are at the center of the predicted $1$:$1$ resonance zone for the chosen cavity frequencies (indicated in Fig.~\ref{fig:energy_13_11_6}(c) by blue lines). Note that in every case the cavity is in the ground state and thus the total energy is below the dissociation energy.  Consistent with our analysis above, and despite a fairly strong coupling value of $\lambda_{c}=0.1$ au, in case of the initial state $|11,0\rangle$ with $\omega_{c} = 2187$ cm$^{-1}$ there is almost no energy exchange between the cavity and the molecule. However, for the other two initial states the presence of the $1$:$1$ resonance leads to cavity-molecule IVR to varying extents depending on the width of the resonance. 

\section{\label{sec:disc} Conclusion and Outlook}%

Our study of the dissociation dynamics of a diatomic molecule coupled to a single cavity mode has revealed several interesting features. Firstly, a significant  suppression of the dissociation probability is observed for cavity frequencies far lower than the fundamental transition frequency of the diatomic molecule. Secondly, the suppression is absent within the linear dipole approximation. Therefore, one has to be careful with theoretical analysis based on the linear dipole approximation. Thirdly, the suppression is also observed in the classical dynamical results. Consequently, the mechanism of suppression could be traced back to features in the classical phase space  giving rise to localization in the phase space. In turn, our analysis reveals that the reduced dissociation is in part due to inefficient cavity-molecule IVR from certain vibrational states of the molecule.  Finally, no significant effect of the cavity molecule coupling on the dissociation dynamics is observed when the cavity frequency is resonant with the fundamental transition of the diatomic molecule. On the other hand, an earlier study by Triana and Herrera \cite{triana2020self}, which neglects the DSE, has argued that enhancement of dissociation probabilities, relative to the cavity free case, is possible for $\omega_{c} \approx \omega_{01}$ provided the coupling $\lambda_{c}$ is sufficiently large with the initial state corresponding to both the diatomic molecule and the cavity being in their ground states.  From the perspective of the current work this suggests that the coupling regime, the nature of the initial state, and the extent of anharmonicity are all crucial in understanding the influence of the cavity on the bond breaking process. However, a detailed understanding of the Triana-Herrera regime in terms of the analysis presented here is worth undertaking.

It is important to note that there are no barriers in the configuration space for the Morse oscillator model. Thus, it is not readily apparent to us that one can apply the theories that invoke ``resonance" between barrier frequencies and the cavity frequency. Of course, as has been established decades ago, barriers are present in the phase space and a proper formulation of TST involves such dynamical bottlenecks \cite{wigner1938transition,waalkens2007wigner,uzertstadvchemphys,zhao2005classical}. Indeed, the present work highlights the role of classical phase space towards understanding polariton chemistry in the VSC regime.

Our focus in this work has been on a rather simple model system and it remains to be seen if our observations can be extended to multiple molecules coupled to the cavity. In particular, the assumption of a single cavity mode is not realistic in the experimental context and the importance of going beyond single mode cavity descriptions in the collective regime  has been highlighted \cite{ribeiro2022multimode,vurgaftman2022comparative,zhou2022interplay}. Nevertheless, in terms of a qualitative understanding of polariton chemistry, our study suggests that for polyatomic molecules in a cavity one has to carefully consider several aspects of IVR and its influence on reactivity,  as also emphasized by several recent studies \cite{schafer2021shining,li2022energy,sidler2022perspective,sun2022suppression,fischer2022cavity}. For example, even in a triatomic molecule there are three vibrational modes and the classical dissociation dynamics is sensitive to the intricate structures \cite{ASKSmolphys,BorondoHCN,SKPKYKS2020} formed by the various nonlinear resonances involving the three modes. There is also evidence that the quantum IVR dynamics is significantly influenced  by this network of classical nonlinear resonances \cite{Semparithi2006,ManikandanKS}. Consequently, tuning the cavity frequency can result in an interesting competition between mode-mode and cavity-mode IVR processes in polyatomic molecules. Thus, it is essential to understand polaritonic reaction dynamics in terms of the role of the shape of the dipole function in different mode directions combined with the modulation of the intramolecular energy flow pathways due to VSC. The recent studies \cite{wang2022chemical,wang2022ivr} by Wang \textit{et al}. on cavity-mediated unimolecular reaction illustrate several interesting regimes of energy exchange that can occur in multidimensional systems. Given the current level of understanding of IVR both from the phase space \cite{karmakar2020intramolecular,farantos2009energy} and quantum \cite{leitner2015quantum,gruebele2004vibrational} perspectives, one can certainly hope for identifying a key mechanistic component of polariton chemistry.

\begin{acknowledgments}
 SK is  grateful to Jino George and Anoop Thomas for discussions. SM is grateful to the Ministry of Education, Government of India for the Prime Minister Research Fellowship (PMRF). SM acknowledges the IIT Kanpur High Performance Computing center for providing computing resources.
\end{acknowledgments}

\appendix

\section{Methodology}
\subsection{Quantum dissociation probabilities}

Starting with the initial polarition wavepacket, we solve the corresponding time-dependent Schr\"{o}dinger equation on a grid using the well established split-operator method \cite{feit1983solution} involving the short time propagator
\begin{equation}
    \hat{U}(\Delta t) = \exp\left(-i\frac{\Delta t}{2\hbar}\hat{V}\right) \exp \left(-i\frac{\Delta t}{\hbar}\hat{T} \right) \exp \left(-i\frac{\Delta t}{2\hbar}\hat{V}\right) \label{split_opr}
\end{equation}
with $\hat{T}$ and $\hat{V}$ being the kinetic and potential energy operators respectively. The value of $\hbar$ is taken as unity in atomic unit. Initial states are time evolved over the timescale of interest ($\sim 5$ ps) with a step size of $\Delta t = 4$ au to ensure convergence of the dissociation probabilities. For the calculations reported here we chose a $1024 \times 1024$ spatial grid. The range of spatial coordinates for the molecule and the cavity are $[0.0, 16.74]$ and $[-300, 300]$ respectively. An optical potential has been employed to avoid unphysical reflection at the grid boundaries \cite{brown1986barriers,leforestier1983optical}.
\begin{equation}
    V_{\rm opt}(q) = -\frac{iV_{0}}{1+e^{[-(q-q^{*})/\eta^{\prime}]}} \label{opt_potential}
\end{equation}
with parameters (in atomic units) $V_{0} = 0.02$, $q^{*} = 16.74$ and $\eta^{\prime} = 0.35$. The introduction of $V_{\rm opt}$ smoothly damps the outgoing wave function.

\subsection{Classical dissociation probabilities}

The classical analog of the wave packet is a Gaussian probability distribution $\rho(\mathbf{Q},\mathbf{P})$ with the same widths in position and momentum. Here, $\mathbf{Q}=(q,q_{c})$ and the corresponding conjugate momenta are denoted as $\mathbf{P}=(p,p_{c})$. The time evolution of this distribution is given by
\begin{equation}
\begin{split}
    \rho_{cl}(\mathbf{Q},\mathbf{P},t) & = \int\int d\mathbf{Q}^{\prime}d\mathbf{P}^{\prime} \delta[\mathbf{Q}-\mathbf{Q}_{t}(\mathbf{P}^{\prime},\mathbf{Q}^{\prime})] \\ & \times \delta[\mathbf{P}-\mathbf{P}_{t}(\mathbf{P}^{\prime},\mathbf{Q}^{\prime})] \rho_{cl}(\mathbf{Q}^{\prime},\mathbf{P}^{\prime},0)  \label{rho_cl_xpx_t}
\end{split}
\end{equation}
where $\mathbf{Q}_{t}(\mathbf{P}^{\prime},\mathbf{Q}^{\prime})$ and $\mathbf{P}_{t}(\mathbf{P}^{\prime},\mathbf{Q}^{\prime})$ represent the time-evolved phase-space coordinates as functions of the initial conditions. Since, our initial wavefunction is a product of a displaced wavepacket and the ground state of the cavity mode, the corresponding classical distribution $\rho_{cl}(\mathbf{Q},\mathbf{P},0) = \rho_{cl}(q,p,0) \rho_{cl}(q_{c},p_{c},0) \label{rho_cl_tot_0}$ is a Wigner distribution centered at $(q_{0}, p_{0},q_{c_{0}},p_{c_{0}})$ in the phase space.
The form of $\rho_{cl}(q,p,0)$ is given by the expression
\begin{equation}
        \rho_{cl}(q,p,0)  = \left(\frac{1}{\pi \sigma_{q} \sigma_{p}} \right)  \exp \left[-\frac{(q-q_{0})^{2}}{\sigma^{2}_{q}} - \frac{p^{2}}{\sigma^{2}_{p}}\right]
\end{equation}
In the above, the position and momentum widths are denoted by $\sigma_{q} = (\hbar/m\omega_{0})^{1/2}$ and $\sigma_{p} = (\hbar m\omega_{0})^{1/2}$ respectively. Similarly, for the cavity mode we have
\begin{equation}
    \rho_{cl}(q_{c},p_{c},0) = \left(\frac{1}{\pi \sigma_{q_{c}} \sigma_{p_{c}}} \right)  \exp \left[-\frac{q_{c}^{2}}{\sigma^{2}_{q_{c}}} - \frac{p_{c}^{2}}{\sigma^{2}_{p_{c}}}\right] 
\end{equation}
In Eqn.~\ref{rho_cl_xpx_t}, formal solution of the Lioville equation, $\left(\mathbf{Q}_{t}, \mathbf{P}_{t}\right)$ is the classical trajectory with the initial condition $\left(\mathbf{Q}^{\prime}, \mathbf{P}^{\prime}\right)$. The classical density at $t=0$ is chosen as Gaussians in the phase space with position and momentum widths consistent with the initial quantum wavepacket density $|\Psi(\mathbf{Q},0)|^{2}$. In order to compute $P^{CM}_{D}(t)$, we initiate an ensemble of $50000$ initial conditions sampled according to the initial density $\rho_{cl}(\mathbf{Q},\mathbf{P},0)$ and integrate their equation of motion forward in time for $5$ ps using a fourth-order Runge – Kutta method.

%\nocite{*}

\bibliography{HFcavity.bib}% Produces the bibliography via BibTeX.

\begin{thebibliography}{77}%
\makeatletter
\providecommand \@ifxundefined [1]{%
 \@ifx{#1\undefined}
}%
\providecommand \@ifnum [1]{%
 \ifnum #1\expandafter \@firstoftwo
 \else \expandafter \@secondoftwo
 \fi
}%
\providecommand \@ifx [1]{%
 \ifx #1\expandafter \@firstoftwo
 \else \expandafter \@secondoftwo
 \fi
}%
\providecommand \natexlab [1]{#1}%
\providecommand \enquote  [1]{``#1''}%
\providecommand \bibnamefont  [1]{#1}%
\providecommand \bibfnamefont [1]{#1}%
\providecommand \citenamefont [1]{#1}%
\providecommand \href@noop [0]{\@secondoftwo}%
\providecommand \href [0]{\begingroup \@sanitize@url \@href}%
\providecommand \@href[1]{\@@startlink{#1}\@@href}%
\providecommand \@@href[1]{\endgroup#1\@@endlink}%
\providecommand \@sanitize@url [0]{\catcode `\\12\catcode `\$12\catcode
  `\&12\catcode `\#12\catcode `\^12\catcode `\_12\catcode `\%12\relax}%
\providecommand \@@startlink[1]{}%
\providecommand \@@endlink[0]{}%
\providecommand \url  [0]{\begingroup\@sanitize@url \@url }%
\providecommand \@url [1]{\endgroup\@href {#1}{\urlprefix }}%
\providecommand \urlprefix  [0]{URL }%
\providecommand \Eprint [0]{\href }%
\providecommand \doibase [0]{https://doi.org/}%
\providecommand \selectlanguage [0]{\@gobble}%
\providecommand \bibinfo  [0]{\@secondoftwo}%
\providecommand \bibfield  [0]{\@secondoftwo}%
\providecommand \translation [1]{[#1]}%
\providecommand \BibitemOpen [0]{}%
\providecommand \bibitemStop [0]{}%
\providecommand \bibitemNoStop [0]{.\EOS\space}%
\providecommand \EOS [0]{\spacefactor3000\relax}%
\providecommand \BibitemShut  [1]{\csname bibitem#1\endcsname}%
\let\auto@bib@innerbib\@empty
%</preamble>
\bibitem [{\citenamefont {Thomas}\ \emph {et~al.}(2016)\citenamefont {Thomas},
  \citenamefont {George}, \citenamefont {Shalabney}, \citenamefont {Dryzhakov},
  \citenamefont {Varma}, \citenamefont {Moran}, \citenamefont {Chervy},
  \citenamefont {Zhong}, \citenamefont {Devaux}, \citenamefont {Genet} \emph
  {et~al.}}]{thomas2016ground}%
  \BibitemOpen
  \bibfield  {author} {\bibinfo {author} {\bibfnamefont {A.}~\bibnamefont
  {Thomas}}, \bibinfo {author} {\bibfnamefont {J.}~\bibnamefont {George}},
  \bibinfo {author} {\bibfnamefont {A.}~\bibnamefont {Shalabney}}, \bibinfo
  {author} {\bibfnamefont {M.}~\bibnamefont {Dryzhakov}}, \bibinfo {author}
  {\bibfnamefont {S.~J.}\ \bibnamefont {Varma}}, \bibinfo {author}
  {\bibfnamefont {J.}~\bibnamefont {Moran}}, \bibinfo {author} {\bibfnamefont
  {T.}~\bibnamefont {Chervy}}, \bibinfo {author} {\bibfnamefont
  {X.}~\bibnamefont {Zhong}}, \bibinfo {author} {\bibfnamefont
  {E.}~\bibnamefont {Devaux}}, \bibinfo {author} {\bibfnamefont
  {C.}~\bibnamefont {Genet}}, \emph {et~al.},\ }\bibfield  {title} {\bibinfo
  {title} {Ground-state chemical reactivity under vibrational coupling to the
  vacuum electromagnetic field},\ }\href@noop {} {\bibfield  {journal}
  {\bibinfo  {journal} {Angew. Chem. Int. Ed.}\ }\textbf {\bibinfo {volume}
  {128}},\ \bibinfo {pages} {11634} (\bibinfo {year} {2016})}\BibitemShut
  {NoStop}%
\bibitem [{\citenamefont {Thomas}\ \emph {et~al.}(2019)\citenamefont {Thomas},
  \citenamefont {Lethuillier-Karl}, \citenamefont {Nagarajan}, \citenamefont
  {Vergauwe}, \citenamefont {George}, \citenamefont {Chervy}, \citenamefont
  {Shalabney}, \citenamefont {Devaux}, \citenamefont {Genet}, \citenamefont
  {Moran} \emph {et~al.}}]{thomas2019tilting}%
  \BibitemOpen
  \bibfield  {author} {\bibinfo {author} {\bibfnamefont {A.}~\bibnamefont
  {Thomas}}, \bibinfo {author} {\bibfnamefont {L.}~\bibnamefont
  {Lethuillier-Karl}}, \bibinfo {author} {\bibfnamefont {K.}~\bibnamefont
  {Nagarajan}}, \bibinfo {author} {\bibfnamefont {R.~M.}\ \bibnamefont
  {Vergauwe}}, \bibinfo {author} {\bibfnamefont {J.}~\bibnamefont {George}},
  \bibinfo {author} {\bibfnamefont {T.}~\bibnamefont {Chervy}}, \bibinfo
  {author} {\bibfnamefont {A.}~\bibnamefont {Shalabney}}, \bibinfo {author}
  {\bibfnamefont {E.}~\bibnamefont {Devaux}}, \bibinfo {author} {\bibfnamefont
  {C.}~\bibnamefont {Genet}}, \bibinfo {author} {\bibfnamefont
  {J.}~\bibnamefont {Moran}}, \emph {et~al.},\ }\bibfield  {title} {\bibinfo
  {title} {Tilting a ground-state reactivity landscape by vibrational strong
  coupling},\ }\href@noop {} {\bibfield  {journal} {\bibinfo  {journal}
  {Science}\ }\textbf {\bibinfo {volume} {363}},\ \bibinfo {pages} {615}
  (\bibinfo {year} {2019})}\BibitemShut {NoStop}%
\bibitem [{\citenamefont {Lather}\ \emph {et~al.}(2019)\citenamefont {Lather},
  \citenamefont {Bhatt}, \citenamefont {Thomas}, \citenamefont {Ebbesen},\ and\
  \citenamefont {George}}]{lather2019cavity}%
  \BibitemOpen
  \bibfield  {author} {\bibinfo {author} {\bibfnamefont {J.}~\bibnamefont
  {Lather}}, \bibinfo {author} {\bibfnamefont {P.}~\bibnamefont {Bhatt}},
  \bibinfo {author} {\bibfnamefont {A.}~\bibnamefont {Thomas}}, \bibinfo
  {author} {\bibfnamefont {T.~W.}\ \bibnamefont {Ebbesen}},\ and\ \bibinfo
  {author} {\bibfnamefont {J.}~\bibnamefont {George}},\ }\bibfield  {title}
  {\bibinfo {title} {Cavity catalysis by cooperative vibrational strong
  coupling of reactant and solvent molecules},\ }\href@noop {} {\bibfield
  {journal} {\bibinfo  {journal} {Angew. Chem. Int. Ed.}\ }\textbf {\bibinfo
  {volume} {131}},\ \bibinfo {pages} {10745} (\bibinfo {year}
  {2019})}\BibitemShut {NoStop}%
\bibitem [{\citenamefont {Vergauwe}\ \emph {et~al.}(2019)\citenamefont
  {Vergauwe}, \citenamefont {Thomas}, \citenamefont {Nagarajan}, \citenamefont
  {Shalabney}, \citenamefont {George}, \citenamefont {Chervy}, \citenamefont
  {Seidel}, \citenamefont {Devaux}, \citenamefont {Torbeev},\ and\
  \citenamefont {Ebbesen}}]{vergauwe2019modification}%
  \BibitemOpen
  \bibfield  {author} {\bibinfo {author} {\bibfnamefont {R.~M.}\ \bibnamefont
  {Vergauwe}}, \bibinfo {author} {\bibfnamefont {A.}~\bibnamefont {Thomas}},
  \bibinfo {author} {\bibfnamefont {K.}~\bibnamefont {Nagarajan}}, \bibinfo
  {author} {\bibfnamefont {A.}~\bibnamefont {Shalabney}}, \bibinfo {author}
  {\bibfnamefont {J.}~\bibnamefont {George}}, \bibinfo {author} {\bibfnamefont
  {T.}~\bibnamefont {Chervy}}, \bibinfo {author} {\bibfnamefont
  {M.}~\bibnamefont {Seidel}}, \bibinfo {author} {\bibfnamefont
  {E.}~\bibnamefont {Devaux}}, \bibinfo {author} {\bibfnamefont
  {V.}~\bibnamefont {Torbeev}},\ and\ \bibinfo {author} {\bibfnamefont {T.~W.}\
  \bibnamefont {Ebbesen}},\ }\bibfield  {title} {\bibinfo {title} {Modification
  of enzyme activity by vibrational strong coupling of water},\ }\href@noop {}
  {\bibfield  {journal} {\bibinfo  {journal} {Angew. Chem. Int. Ed.}\ }\textbf
  {\bibinfo {volume} {58}},\ \bibinfo {pages} {15324} (\bibinfo {year}
  {2019})}\BibitemShut {NoStop}%
\bibitem [{\citenamefont {Thomas}\ \emph {et~al.}(2020)\citenamefont {Thomas},
  \citenamefont {Jayachandran}, \citenamefont {Lethuillier-Karl}, \citenamefont
  {Vergauwe}, \citenamefont {Nagarajan}, \citenamefont {Devaux}, \citenamefont
  {Genet}, \citenamefont {Moran},\ and\ \citenamefont
  {Ebbesen}}]{thomas2020ground}%
  \BibitemOpen
  \bibfield  {author} {\bibinfo {author} {\bibfnamefont {A.}~\bibnamefont
  {Thomas}}, \bibinfo {author} {\bibfnamefont {A.}~\bibnamefont
  {Jayachandran}}, \bibinfo {author} {\bibfnamefont {L.}~\bibnamefont
  {Lethuillier-Karl}}, \bibinfo {author} {\bibfnamefont {R.~M.}\ \bibnamefont
  {Vergauwe}}, \bibinfo {author} {\bibfnamefont {K.}~\bibnamefont {Nagarajan}},
  \bibinfo {author} {\bibfnamefont {E.}~\bibnamefont {Devaux}}, \bibinfo
  {author} {\bibfnamefont {C.}~\bibnamefont {Genet}}, \bibinfo {author}
  {\bibfnamefont {J.}~\bibnamefont {Moran}},\ and\ \bibinfo {author}
  {\bibfnamefont {T.~W.}\ \bibnamefont {Ebbesen}},\ }\bibfield  {title}
  {\bibinfo {title} {Ground state chemistry under vibrational strong coupling:
  dependence of thermodynamic parameters on the {R}abi splitting energy},\
  }\href@noop {} {\bibfield  {journal} {\bibinfo  {journal} {Nanophotonics}\
  }\textbf {\bibinfo {volume} {9}},\ \bibinfo {pages} {249} (\bibinfo {year}
  {2020})}\BibitemShut {NoStop}%
\bibitem [{\citenamefont {Nagarajan}\ \emph {et~al.}(2021)\citenamefont
  {Nagarajan}, \citenamefont {Thomas},\ and\ \citenamefont
  {Ebbesen}}]{nagarajan2021chemistry}%
  \BibitemOpen
  \bibfield  {author} {\bibinfo {author} {\bibfnamefont {K.}~\bibnamefont
  {Nagarajan}}, \bibinfo {author} {\bibfnamefont {A.}~\bibnamefont {Thomas}},\
  and\ \bibinfo {author} {\bibfnamefont {T.~W.}\ \bibnamefont {Ebbesen}},\
  }\bibfield  {title} {\bibinfo {title} {Chemistry under vibrational strong
  coupling},\ }\href@noop {} {\bibfield  {journal} {\bibinfo  {journal} {J. Am.
  Chem. Soc.}\ }\textbf {\bibinfo {volume} {143}},\ \bibinfo {pages} {16877}
  (\bibinfo {year} {2021})}\BibitemShut {NoStop}%
\bibitem [{\citenamefont {Haroche}\ and\ \citenamefont
  {Raimond}(2006)}]{haroche2006exploring}%
  \BibitemOpen
  \bibfield  {author} {\bibinfo {author} {\bibfnamefont {S.}~\bibnamefont
  {Haroche}}\ and\ \bibinfo {author} {\bibfnamefont {J.~M.}\ \bibnamefont
  {Raimond}},\ }\href {https://books.google.co.in/books?id=ynwSDAAAQBAJ} {\emph
  {\bibinfo {title} {Exploring the Quantum: Atoms, Cavities, and Photons}}}\
  (\bibinfo  {publisher} {Oxford university press},\ \bibinfo {year}
  {2006})\BibitemShut {NoStop}%
\bibitem [{\citenamefont {Bloembergen}\ and\ \citenamefont
  {Zewail}(1984)}]{bloembergen1984energy}%
  \BibitemOpen
  \bibfield  {author} {\bibinfo {author} {\bibfnamefont {N.}~\bibnamefont
  {Bloembergen}}\ and\ \bibinfo {author} {\bibfnamefont {A.~H.}\ \bibnamefont
  {Zewail}},\ }\bibfield  {title} {\bibinfo {title} {Energy redistribution in
  isolated molecules and the question of mode-selective laser chemistry
  revisited},\ }\href@noop {} {\bibfield  {journal} {\bibinfo  {journal} {J.
  Phys. Chem.}\ }\textbf {\bibinfo {volume} {88}},\ \bibinfo {pages} {5459}
  (\bibinfo {year} {1984})}\BibitemShut {NoStop}%
\bibitem [{\citenamefont {Nesbitt}\ and\ \citenamefont
  {Field}(1996)}]{nesbitt1996vibrational}%
  \BibitemOpen
  \bibfield  {author} {\bibinfo {author} {\bibfnamefont {D.~J.}\ \bibnamefont
  {Nesbitt}}\ and\ \bibinfo {author} {\bibfnamefont {R.~W.}\ \bibnamefont
  {Field}},\ }\bibfield  {title} {\bibinfo {title} {Vibrational energy flow in
  highly excited molecules: Role of intramolecular vibrational
  redistribution},\ }\href@noop {} {\bibfield  {journal} {\bibinfo  {journal}
  {J. Phys. Chem.}\ }\textbf {\bibinfo {volume} {100}},\ \bibinfo {pages}
  {12735} (\bibinfo {year} {1996})}\BibitemShut {NoStop}%
\bibitem [{\citenamefont {Uzer}\ and\ \citenamefont
  {Miller}(1991)}]{uzermillerphysrep}%
  \BibitemOpen
  \bibfield  {author} {\bibinfo {author} {\bibfnamefont {T.}~\bibnamefont
  {Uzer}}\ and\ \bibinfo {author} {\bibfnamefont {W.}~\bibnamefont {Miller}},\
  }\bibfield  {title} {\bibinfo {title} {Theories of intramolecular vibrational
  energy transfer},\ }\href
  {https://doi.org/https://doi.org/10.1016/0370-1573(91)90140-H} {\bibfield
  {journal} {\bibinfo  {journal} {Phys. Rep.}\ }\textbf {\bibinfo {volume}
  {199}},\ \bibinfo {pages} {73} (\bibinfo {year} {1991})}\BibitemShut
  {NoStop}%
\bibitem [{\citenamefont {Sch{\"a}fer}\ \emph {et~al.}(2021)\citenamefont
  {Sch{\"a}fer}, \citenamefont {Flick}, \citenamefont {Ronca}, \citenamefont
  {Narang},\ and\ \citenamefont {Rubio}}]{schafer2021shining}%
  \BibitemOpen
  \bibfield  {author} {\bibinfo {author} {\bibfnamefont {C.}~\bibnamefont
  {Sch{\"a}fer}}, \bibinfo {author} {\bibfnamefont {J.}~\bibnamefont {Flick}},
  \bibinfo {author} {\bibfnamefont {E.}~\bibnamefont {Ronca}}, \bibinfo
  {author} {\bibfnamefont {P.}~\bibnamefont {Narang}},\ and\ \bibinfo {author}
  {\bibfnamefont {A.}~\bibnamefont {Rubio}},\ }\bibfield  {title} {\bibinfo
  {title} {Shining light on the microscopic resonant mechanism responsible for
  cavity-mediated chemical reactivity},\ }\href@noop {} {\bibfield  {journal}
  {\bibinfo  {journal} {arXiv:2104.12429}\ } (\bibinfo {year}
  {2021})}\BibitemShut {NoStop}%
\bibitem [{\citenamefont {Chen}\ \emph {et~al.}(2022)\citenamefont {Chen},
  \citenamefont {Du}, \citenamefont {Yang}, \citenamefont {Yuen-Zhou},\ and\
  \citenamefont {Xiong}}]{chen2022cavity}%
  \BibitemOpen
  \bibfield  {author} {\bibinfo {author} {\bibfnamefont {T.~T.}\ \bibnamefont
  {Chen}}, \bibinfo {author} {\bibfnamefont {M.}~\bibnamefont {Du}}, \bibinfo
  {author} {\bibfnamefont {Z.}~\bibnamefont {Yang}}, \bibinfo {author}
  {\bibfnamefont {J.}~\bibnamefont {Yuen-Zhou}},\ and\ \bibinfo {author}
  {\bibfnamefont {W.}~\bibnamefont {Xiong}},\ }\bibfield  {title} {\bibinfo
  {title} {Cavity-enabled enhancement of ultrafast intramolecular vibrational
  redistribution over pseudorotation},\ }\href@noop {} {\bibfield  {journal}
  {\bibinfo  {journal} {Science}\ }\textbf {\bibinfo {volume} {378}},\ \bibinfo
  {pages} {790} (\bibinfo {year} {2022})}\BibitemShut {NoStop}%
\bibitem [{\citenamefont {Wang}\ \emph
  {et~al.}(2022{\natexlab{a}})\citenamefont {Wang}, \citenamefont {Flick},\
  and\ \citenamefont {Yelin}}]{wang2022chemical}%
  \BibitemOpen
  \bibfield  {author} {\bibinfo {author} {\bibfnamefont {D.~S.}\ \bibnamefont
  {Wang}}, \bibinfo {author} {\bibfnamefont {J.}~\bibnamefont {Flick}},\ and\
  \bibinfo {author} {\bibfnamefont {S.~F.}\ \bibnamefont {Yelin}},\ }\bibfield
  {title} {\bibinfo {title} {Chemical reactivity under collective vibrational
  strong coupling},\ }\href@noop {} {\bibfield  {journal} {\bibinfo  {journal}
  {J. Chem. Phys.}\ }\textbf {\bibinfo {volume} {157}},\ \bibinfo {pages}
  {224304} (\bibinfo {year} {2022}{\natexlab{a}})}\BibitemShut {NoStop}%
\bibitem [{\citenamefont {Hern{\'a}ndez}\ and\ \citenamefont
  {Herrera}(2019)}]{hernandez2019multi}%
  \BibitemOpen
  \bibfield  {author} {\bibinfo {author} {\bibfnamefont {F.~J.}\ \bibnamefont
  {Hern{\'a}ndez}}\ and\ \bibinfo {author} {\bibfnamefont {F.}~\bibnamefont
  {Herrera}},\ }\bibfield  {title} {\bibinfo {title} {Multi-level quantum
  {R}abi model for anharmonic vibrational polaritons},\ }\href@noop {}
  {\bibfield  {journal} {\bibinfo  {journal} {J. Chem. Phys.}\ }\textbf
  {\bibinfo {volume} {151}},\ \bibinfo {pages} {144116} (\bibinfo {year}
  {2019})}\BibitemShut {NoStop}%
\bibitem [{\citenamefont {Gruebele}\ and\ \citenamefont
  {Wolynes}(2004)}]{gruebele2004vibrational}%
  \BibitemOpen
  \bibfield  {author} {\bibinfo {author} {\bibfnamefont {M.}~\bibnamefont
  {Gruebele}}\ and\ \bibinfo {author} {\bibfnamefont {P.~G.}\ \bibnamefont
  {Wolynes}},\ }\bibfield  {title} {\bibinfo {title} {Vibrational energy flow
  and chemical reactions},\ }\href@noop {} {\bibfield  {journal} {\bibinfo
  {journal} {Acc. Chem. Res.}\ }\textbf {\bibinfo {volume} {37}},\ \bibinfo
  {pages} {261} (\bibinfo {year} {2004})}\BibitemShut {NoStop}%
\bibitem [{\citenamefont {Leitner}(2015)}]{leitner2015quantum}%
  \BibitemOpen
  \bibfield  {author} {\bibinfo {author} {\bibfnamefont {D.~M.}\ \bibnamefont
  {Leitner}},\ }\bibfield  {title} {\bibinfo {title} {Quantum ergodicity and
  energy flow in molecules},\ }\href@noop {} {\bibfield  {journal} {\bibinfo
  {journal} {Adv. Phys.}\ }\textbf {\bibinfo {volume} {64}},\ \bibinfo {pages}
  {445} (\bibinfo {year} {2015})}\BibitemShut {NoStop}%
\bibitem [{\citenamefont {Karmakar}\ and\ \citenamefont
  {Keshavamurthy}(2020)}]{karmakar2020intramolecular}%
  \BibitemOpen
  \bibfield  {author} {\bibinfo {author} {\bibfnamefont {S.}~\bibnamefont
  {Karmakar}}\ and\ \bibinfo {author} {\bibfnamefont {S.}~\bibnamefont
  {Keshavamurthy}},\ }\bibfield  {title} {\bibinfo {title} {Intramolecular
  vibrational energy redistribution and the quantum ergodicity transition: a
  phase space perspective},\ }\href@noop {} {\bibfield  {journal} {\bibinfo
  {journal} {Phys. Chem. Chem. Phys.}\ }\textbf {\bibinfo {volume} {22}},\
  \bibinfo {pages} {11139} (\bibinfo {year} {2020})}\BibitemShut {NoStop}%
\bibitem [{\citenamefont {Farantos}\ \emph {et~al.}(2009)\citenamefont
  {Farantos}, \citenamefont {Schinke}, \citenamefont {Guo},\ and\ \citenamefont
  {Joyeux}}]{farantos2009energy}%
  \BibitemOpen
  \bibfield  {author} {\bibinfo {author} {\bibfnamefont {S.~C.}\ \bibnamefont
  {Farantos}}, \bibinfo {author} {\bibfnamefont {R.}~\bibnamefont {Schinke}},
  \bibinfo {author} {\bibfnamefont {H.}~\bibnamefont {Guo}},\ and\ \bibinfo
  {author} {\bibfnamefont {M.}~\bibnamefont {Joyeux}},\ }\bibfield  {title}
  {\bibinfo {title} {Energy localization in molecules, bifurcation phenomena,
  and their spectroscopic signatures: The global view},\ }\href@noop {}
  {\bibfield  {journal} {\bibinfo  {journal} {Chem. Rev.}\ }\textbf {\bibinfo
  {volume} {109}},\ \bibinfo {pages} {4248} (\bibinfo {year}
  {2009})}\BibitemShut {NoStop}%
\bibitem [{\citenamefont {Campos-Gonzalez-Angulo}\ \emph
  {et~al.}(2019)\citenamefont {Campos-Gonzalez-Angulo}, \citenamefont
  {Ribeiro},\ and\ \citenamefont {Yuen-Zhou}}]{campos2019resonant}%
  \BibitemOpen
  \bibfield  {author} {\bibinfo {author} {\bibfnamefont {J.~A.}\ \bibnamefont
  {Campos-Gonzalez-Angulo}}, \bibinfo {author} {\bibfnamefont {R.~F.}\
  \bibnamefont {Ribeiro}},\ and\ \bibinfo {author} {\bibfnamefont
  {J.}~\bibnamefont {Yuen-Zhou}},\ }\bibfield  {title} {\bibinfo {title}
  {Resonant catalysis of thermally activated chemical reactions with
  vibrational polaritons},\ }\href@noop {} {\bibfield  {journal} {\bibinfo
  {journal} {Nat. Commun.}\ }\textbf {\bibinfo {volume} {10}},\ \bibinfo
  {pages} {1} (\bibinfo {year} {2019})}\BibitemShut {NoStop}%
\bibitem [{\citenamefont {Li}\ \emph {et~al.}(2021{\natexlab{a}})\citenamefont
  {Li}, \citenamefont {Mandal},\ and\ \citenamefont {Huo}}]{li2021cavity}%
  \BibitemOpen
  \bibfield  {author} {\bibinfo {author} {\bibfnamefont {X.}~\bibnamefont
  {Li}}, \bibinfo {author} {\bibfnamefont {A.}~\bibnamefont {Mandal}},\ and\
  \bibinfo {author} {\bibfnamefont {P.}~\bibnamefont {Huo}},\ }\bibfield
  {title} {\bibinfo {title} {Cavity frequency-dependent theory for vibrational
  polariton chemistry},\ }\href@noop {} {\bibfield  {journal} {\bibinfo
  {journal} {Nat. Commun.}\ }\textbf {\bibinfo {volume} {12}},\ \bibinfo
  {pages} {1} (\bibinfo {year} {2021}{\natexlab{a}})}\BibitemShut {NoStop}%
\bibitem [{\citenamefont {Li}\ \emph {et~al.}(2021{\natexlab{b}})\citenamefont
  {Li}, \citenamefont {Mandal},\ and\ \citenamefont {Huo}}]{li2021theory}%
  \BibitemOpen
  \bibfield  {author} {\bibinfo {author} {\bibfnamefont {X.}~\bibnamefont
  {Li}}, \bibinfo {author} {\bibfnamefont {A.}~\bibnamefont {Mandal}},\ and\
  \bibinfo {author} {\bibfnamefont {P.}~\bibnamefont {Huo}},\ }\bibfield
  {title} {\bibinfo {title} {Theory of mode-selective chemistry through
  polaritonic vibrational strong coupling},\ }\href@noop {} {\bibfield
  {journal} {\bibinfo  {journal} {J. Chem. Phys. Lett.}\ }\textbf {\bibinfo
  {volume} {12}},\ \bibinfo {pages} {6974} (\bibinfo {year}
  {2021}{\natexlab{b}})}\BibitemShut {NoStop}%
\bibitem [{\citenamefont {Du}\ and\ \citenamefont
  {Yuen-Zhou}(2022)}]{du2022catalysis}%
  \BibitemOpen
  \bibfield  {author} {\bibinfo {author} {\bibfnamefont {M.}~\bibnamefont
  {Du}}\ and\ \bibinfo {author} {\bibfnamefont {J.}~\bibnamefont {Yuen-Zhou}},\
  }\bibfield  {title} {\bibinfo {title} {Catalysis by dark states in
  vibropolaritonic chemistry},\ }\href@noop {} {\bibfield  {journal} {\bibinfo
  {journal} {Phys. Rev. Lett.}\ }\textbf {\bibinfo {volume} {128}},\ \bibinfo
  {pages} {096001} (\bibinfo {year} {2022})}\BibitemShut {NoStop}%
\bibitem [{\citenamefont {Mandal}\ \emph
  {et~al.}(2022{\natexlab{a}})\citenamefont {Mandal}, \citenamefont {Li},\ and\
  \citenamefont {Huo}}]{mandal2022theory}%
  \BibitemOpen
  \bibfield  {author} {\bibinfo {author} {\bibfnamefont {A.}~\bibnamefont
  {Mandal}}, \bibinfo {author} {\bibfnamefont {X.}~\bibnamefont {Li}},\ and\
  \bibinfo {author} {\bibfnamefont {P.}~\bibnamefont {Huo}},\ }\bibfield
  {title} {\bibinfo {title} {Theory of vibrational polariton chemistry in the
  collective coupling regime},\ }\href@noop {} {\bibfield  {journal} {\bibinfo
  {journal} {J. Chem. Phys.}\ }\textbf {\bibinfo {volume} {156}},\ \bibinfo
  {pages} {014101} (\bibinfo {year} {2022}{\natexlab{a}})}\BibitemShut
  {NoStop}%
\bibitem [{\citenamefont {Wang}\ \emph
  {et~al.}(2022{\natexlab{b}})\citenamefont {Wang}, \citenamefont {Neuman},
  \citenamefont {Yelin},\ and\ \citenamefont {Flick}}]{wang2022ivr}%
  \BibitemOpen
  \bibfield  {author} {\bibinfo {author} {\bibfnamefont {D.~S.}\ \bibnamefont
  {Wang}}, \bibinfo {author} {\bibfnamefont {T.}~\bibnamefont {Neuman}},
  \bibinfo {author} {\bibfnamefont {S.~F.}\ \bibnamefont {Yelin}},\ and\
  \bibinfo {author} {\bibfnamefont {J.}~\bibnamefont {Flick}},\ }\bibfield
  {title} {\bibinfo {title} {Cavity-modified unimolecular dissociation
  reactions via intramolecular vibrational energy redistribution},\ }\href@noop
  {} {\bibfield  {journal} {\bibinfo  {journal} {J. Phys. Chem. Lett.}\
  }\textbf {\bibinfo {volume} {13}},\ \bibinfo {pages} {3317} (\bibinfo {year}
  {2022}{\natexlab{b}})}\BibitemShut {NoStop}%
\bibitem [{\citenamefont {Yang}\ and\ \citenamefont
  {Cao}(2021)}]{yang2021quantum}%
  \BibitemOpen
  \bibfield  {author} {\bibinfo {author} {\bibfnamefont {P.~Y.}\ \bibnamefont
  {Yang}}\ and\ \bibinfo {author} {\bibfnamefont {J.}~\bibnamefont {Cao}},\
  }\bibfield  {title} {\bibinfo {title} {Quantum effects in chemical reactions
  under polaritonic vibrational strong coupling},\ }\href@noop {} {\bibfield
  {journal} {\bibinfo  {journal} {J. Phys. Chem. Lett.}\ }\textbf {\bibinfo
  {volume} {12}},\ \bibinfo {pages} {9531} (\bibinfo {year}
  {2021})}\BibitemShut {NoStop}%
\bibitem [{\citenamefont {Galego}\ \emph {et~al.}(2019)\citenamefont {Galego},
  \citenamefont {Climent}, \citenamefont {Garcia-Vidal},\ and\ \citenamefont
  {Feist}}]{galego2019cavity}%
  \BibitemOpen
  \bibfield  {author} {\bibinfo {author} {\bibfnamefont {J.}~\bibnamefont
  {Galego}}, \bibinfo {author} {\bibfnamefont {C.}~\bibnamefont {Climent}},
  \bibinfo {author} {\bibfnamefont {F.~J.}\ \bibnamefont {Garcia-Vidal}},\ and\
  \bibinfo {author} {\bibfnamefont {J.}~\bibnamefont {Feist}},\ }\bibfield
  {title} {\bibinfo {title} {Cavity {C}asimir-{P}older forces and their effects
  in ground-state chemical reactivity},\ }\href@noop {} {\bibfield  {journal}
  {\bibinfo  {journal} {Phys. Rev. X}\ }\textbf {\bibinfo {volume} {9}},\
  \bibinfo {pages} {021057} (\bibinfo {year} {2019})}\BibitemShut {NoStop}%
\bibitem [{\citenamefont {Li}\ \emph {et~al.}(2020{\natexlab{a}})\citenamefont
  {Li}, \citenamefont {Nitzan},\ and\ \citenamefont {Subotnik}}]{li2020origin}%
  \BibitemOpen
  \bibfield  {author} {\bibinfo {author} {\bibfnamefont {T.~E.}\ \bibnamefont
  {Li}}, \bibinfo {author} {\bibfnamefont {A.}~\bibnamefont {Nitzan}},\ and\
  \bibinfo {author} {\bibfnamefont {J.~E.}\ \bibnamefont {Subotnik}},\
  }\bibfield  {title} {\bibinfo {title} {On the origin of ground-state
  vacuum-field catalysis: Equilibrium consideration},\ }\href@noop {}
  {\bibfield  {journal} {\bibinfo  {journal} {J. Chem. Phys.}\ }\textbf
  {\bibinfo {volume} {152}},\ \bibinfo {pages} {234107} (\bibinfo {year}
  {2020}{\natexlab{a}})}\BibitemShut {NoStop}%
\bibitem [{\citenamefont {Campos-Gonzalez-Angulo}\ and\ \citenamefont
  {Yuen-Zhou}(2020)}]{campos2020polaritonic}%
  \BibitemOpen
  \bibfield  {author} {\bibinfo {author} {\bibfnamefont {J.~A.}\ \bibnamefont
  {Campos-Gonzalez-Angulo}}\ and\ \bibinfo {author} {\bibfnamefont
  {J.}~\bibnamefont {Yuen-Zhou}},\ }\bibfield  {title} {\bibinfo {title}
  {Polaritonic normal modes in transition state theory},\ }\href@noop {}
  {\bibfield  {journal} {\bibinfo  {journal} {J. Chem. Phys.}\ }\textbf
  {\bibinfo {volume} {152}},\ \bibinfo {pages} {161101} (\bibinfo {year}
  {2020})}\BibitemShut {NoStop}%
\bibitem [{\citenamefont {Zhdanov}(2020)}]{zhdanov2020vacuum}%
  \BibitemOpen
  \bibfield  {author} {\bibinfo {author} {\bibfnamefont {V.~P.}\ \bibnamefont
  {Zhdanov}},\ }\bibfield  {title} {\bibinfo {title} {Vacuum field in a cavity,
  light-mediated vibrational coupling, and chemical reactivity},\ }\href@noop
  {} {\bibfield  {journal} {\bibinfo  {journal} {Chem. Phys.}\ }\textbf
  {\bibinfo {volume} {535}},\ \bibinfo {pages} {110767} (\bibinfo {year}
  {2020})}\BibitemShut {NoStop}%
\bibitem [{\citenamefont {Sun}\ and\ \citenamefont
  {Vendrell}(2022)}]{sun2022suppression}%
  \BibitemOpen
  \bibfield  {author} {\bibinfo {author} {\bibfnamefont {J.}~\bibnamefont
  {Sun}}\ and\ \bibinfo {author} {\bibfnamefont {O.}~\bibnamefont {Vendrell}},\
  }\bibfield  {title} {\bibinfo {title} {Suppression and enhancement of thermal
  chemical rates in a cavity},\ }\href@noop {} {\bibfield  {journal} {\bibinfo
  {journal} {J. Phys. Chem. Lett.}\ }\textbf {\bibinfo {volume} {13}},\
  \bibinfo {pages} {4441} (\bibinfo {year} {2022})}\BibitemShut {NoStop}%
\bibitem [{\citenamefont {Lindoy}\ \emph {et~al.}(2022)\citenamefont {Lindoy},
  \citenamefont {Mandal},\ and\ \citenamefont {Reichman}}]{lindoy2022resonant}%
  \BibitemOpen
  \bibfield  {author} {\bibinfo {author} {\bibfnamefont {L.~P.}\ \bibnamefont
  {Lindoy}}, \bibinfo {author} {\bibfnamefont {A.}~\bibnamefont {Mandal}},\
  and\ \bibinfo {author} {\bibfnamefont {D.~R.}\ \bibnamefont {Reichman}},\
  }\bibfield  {title} {\bibinfo {title} {Resonant cavity modification of
  ground-state chemical kinetics},\ }\href@noop {} {\bibfield  {journal}
  {\bibinfo  {journal} {J. Phys. Chem. Lett.}\ }\textbf {\bibinfo {volume}
  {13}},\ \bibinfo {pages} {6580} (\bibinfo {year} {2022})}\BibitemShut
  {NoStop}%
\bibitem [{\citenamefont {Philbin}\ \emph {et~al.}(2022)\citenamefont
  {Philbin}, \citenamefont {Wang}, \citenamefont {Narang},\ and\ \citenamefont
  {Dou}}]{philbin2022chemical}%
  \BibitemOpen
  \bibfield  {author} {\bibinfo {author} {\bibfnamefont {J.~P.}\ \bibnamefont
  {Philbin}}, \bibinfo {author} {\bibfnamefont {Y.}~\bibnamefont {Wang}},
  \bibinfo {author} {\bibfnamefont {P.}~\bibnamefont {Narang}},\ and\ \bibinfo
  {author} {\bibfnamefont {W.}~\bibnamefont {Dou}},\ }\bibfield  {title}
  {\bibinfo {title} {Chemical reactions in imperfect cavities: Enhancement,
  suppression, and resonance},\ }\href@noop {} {\bibfield  {journal} {\bibinfo
  {journal} {J. Phys. Chem. C}\ }\textbf {\bibinfo {volume} {126}},\ \bibinfo
  {pages} {14908} (\bibinfo {year} {2022})}\BibitemShut {NoStop}%
\bibitem [{\citenamefont {Ribeiro}\ \emph {et~al.}(2018)\citenamefont
  {Ribeiro}, \citenamefont {Mart{\'\i}nez-Mart{\'\i}nez}, \citenamefont {Du},
  \citenamefont {Campos-Gonzalez-Angulo},\ and\ \citenamefont
  {Yuen-Zhou}}]{ribeiro2018polariton}%
  \BibitemOpen
  \bibfield  {author} {\bibinfo {author} {\bibfnamefont {R.~F.}\ \bibnamefont
  {Ribeiro}}, \bibinfo {author} {\bibfnamefont {L.~A.}\ \bibnamefont
  {Mart{\'\i}nez-Mart{\'\i}nez}}, \bibinfo {author} {\bibfnamefont
  {M.}~\bibnamefont {Du}}, \bibinfo {author} {\bibfnamefont {J.}~\bibnamefont
  {Campos-Gonzalez-Angulo}},\ and\ \bibinfo {author} {\bibfnamefont
  {J.}~\bibnamefont {Yuen-Zhou}},\ }\bibfield  {title} {\bibinfo {title}
  {Polariton chemistry: controlling molecular dynamics with optical cavities},\
  }\href@noop {} {\bibfield  {journal} {\bibinfo  {journal} {Chem. Sci.}\
  }\textbf {\bibinfo {volume} {9}},\ \bibinfo {pages} {6325} (\bibinfo {year}
  {2018})}\BibitemShut {NoStop}%
\bibitem [{\citenamefont {Botzung}\ \emph {et~al.}(2020)\citenamefont
  {Botzung}, \citenamefont {Hagenm{\"u}ller}, \citenamefont {Sch{\"u}tz},
  \citenamefont {Dubail}, \citenamefont {Pupillo},\ and\ \citenamefont
  {Schachenmayer}}]{botzung2020dark}%
  \BibitemOpen
  \bibfield  {author} {\bibinfo {author} {\bibfnamefont {T.}~\bibnamefont
  {Botzung}}, \bibinfo {author} {\bibfnamefont {D.}~\bibnamefont
  {Hagenm{\"u}ller}}, \bibinfo {author} {\bibfnamefont {S.}~\bibnamefont
  {Sch{\"u}tz}}, \bibinfo {author} {\bibfnamefont {J.}~\bibnamefont {Dubail}},
  \bibinfo {author} {\bibfnamefont {G.}~\bibnamefont {Pupillo}},\ and\ \bibinfo
  {author} {\bibfnamefont {J.}~\bibnamefont {Schachenmayer}},\ }\bibfield
  {title} {\bibinfo {title} {Dark state semilocalization of quantum emitters in
  a cavity},\ }\href@noop {} {\bibfield  {journal} {\bibinfo  {journal} {Phys.
  Rev. B}\ }\textbf {\bibinfo {volume} {102}},\ \bibinfo {pages} {144202}
  (\bibinfo {year} {2020})}\BibitemShut {NoStop}%
\bibitem [{\citenamefont {Campos-Gonzalez-Angulo}\ and\ \citenamefont
  {Yuen-Zhou}(2022)}]{zhoutaviscummings2022}%
  \BibitemOpen
  \bibfield  {author} {\bibinfo {author} {\bibfnamefont {J.~A.}\ \bibnamefont
  {Campos-Gonzalez-Angulo}}\ and\ \bibinfo {author} {\bibfnamefont
  {J.}~\bibnamefont {Yuen-Zhou}},\ }\bibfield  {title} {\bibinfo {title}
  {Generalization of the {T}avis–{C}ummings model for multi-level anharmonic
  systems: Insights on the second excitation manifold},\ }\href@noop {}
  {\bibfield  {journal} {\bibinfo  {journal} {J. Chem. Phys.}\ }\textbf
  {\bibinfo {volume} {156}},\ \bibinfo {pages} {194308} (\bibinfo {year}
  {2022})}\BibitemShut {NoStop}%
\bibitem [{\citenamefont {Gera}\ and\ \citenamefont
  {Sebastian}(2022)}]{gera2022effects}%
  \BibitemOpen
  \bibfield  {author} {\bibinfo {author} {\bibfnamefont {T.}~\bibnamefont
  {Gera}}\ and\ \bibinfo {author} {\bibfnamefont {K.~L.}\ \bibnamefont
  {Sebastian}},\ }\bibfield  {title} {\bibinfo {title} {Effects of disorder on
  polaritonic and dark states in a cavity using the disordered
  {T}avis--{C}ummings model},\ }\href@noop {} {\bibfield  {journal} {\bibinfo
  {journal} {J. Chem. Phys.}\ }\textbf {\bibinfo {volume} {156}},\ \bibinfo
  {pages} {194304} (\bibinfo {year} {2022})}\BibitemShut {NoStop}%
\bibitem [{\citenamefont {Li}\ \emph {et~al.}(2020{\natexlab{b}})\citenamefont
  {Li}, \citenamefont {Subotnik},\ and\ \citenamefont
  {Nitzan}}]{li2020cavitywater}%
  \BibitemOpen
  \bibfield  {author} {\bibinfo {author} {\bibfnamefont {T.~E.}\ \bibnamefont
  {Li}}, \bibinfo {author} {\bibfnamefont {J.~E.}\ \bibnamefont {Subotnik}},\
  and\ \bibinfo {author} {\bibfnamefont {A.}~\bibnamefont {Nitzan}},\
  }\bibfield  {title} {\bibinfo {title} {Cavity molecular dynamics simulations
  of liquid water under vibrational ultrastrong coupling},\ }\href@noop {}
  {\bibfield  {journal} {\bibinfo  {journal} {Proc. Natl. Acad. Sci.}\ }\textbf
  {\bibinfo {volume} {117}},\ \bibinfo {pages} {18324} (\bibinfo {year}
  {2020}{\natexlab{b}})}\BibitemShut {NoStop}%
\bibitem [{\citenamefont {Li}\ \emph {et~al.}(2021{\natexlab{c}})\citenamefont
  {Li}, \citenamefont {Nitzan},\ and\ \citenamefont
  {Subotnik}}]{li2021collective}%
  \BibitemOpen
  \bibfield  {author} {\bibinfo {author} {\bibfnamefont {T.~E.}\ \bibnamefont
  {Li}}, \bibinfo {author} {\bibfnamefont {A.}~\bibnamefont {Nitzan}},\ and\
  \bibinfo {author} {\bibfnamefont {J.~E.}\ \bibnamefont {Subotnik}},\
  }\bibfield  {title} {\bibinfo {title} {Collective vibrational strong coupling
  effects on molecular vibrational relaxation and energy transfer: Numerical
  insights via cavity molecular dynamics simulations},\ }\href@noop {}
  {\bibfield  {journal} {\bibinfo  {journal} {Angew. Chem. Int. Ed.}\ }\textbf
  {\bibinfo {volume} {133}},\ \bibinfo {pages} {15661} (\bibinfo {year}
  {2021}{\natexlab{c}})}\BibitemShut {NoStop}%
\bibitem [{\citenamefont {Li}\ \emph {et~al.}(2022{\natexlab{a}})\citenamefont
  {Li}, \citenamefont {Nitzan},\ and\ \citenamefont
  {Subotnik}}]{li2022polaritonmd}%
  \BibitemOpen
  \bibfield  {author} {\bibinfo {author} {\bibfnamefont {T.~E.}\ \bibnamefont
  {Li}}, \bibinfo {author} {\bibfnamefont {A.}~\bibnamefont {Nitzan}},\ and\
  \bibinfo {author} {\bibfnamefont {J.~E.}\ \bibnamefont {Subotnik}},\
  }\bibfield  {title} {\bibinfo {title} {{Polariton relaxation under
  vibrational strong coupling: Comparing cavity molecular dynamics simulations
  against Fermi’s golden rule rate}},\ }\href@noop {} {\bibfield  {journal}
  {\bibinfo  {journal} {J. Chem. Phys.}\ }\textbf {\bibinfo {volume} {156}},\
  \bibinfo {pages} {134106} (\bibinfo {year} {2022}{\natexlab{a}})}\BibitemShut
  {NoStop}%
\bibitem [{\citenamefont {Li}\ \emph {et~al.}(2022{\natexlab{b}})\citenamefont
  {Li}, \citenamefont {Nitzan}, \citenamefont {Hammes-Schiffer},\ and\
  \citenamefont {Subotnik}}]{li2022pimd}%
  \BibitemOpen
  \bibfield  {author} {\bibinfo {author} {\bibfnamefont {T.~E.}\ \bibnamefont
  {Li}}, \bibinfo {author} {\bibfnamefont {A.}~\bibnamefont {Nitzan}}, \bibinfo
  {author} {\bibfnamefont {S.}~\bibnamefont {Hammes-Schiffer}},\ and\ \bibinfo
  {author} {\bibfnamefont {J.~E.}\ \bibnamefont {Subotnik}},\ }\bibfield
  {title} {\bibinfo {title} {Quantum simulations of vibrational strong coupling
  via path integrals},\ }\href@noop {} {\bibfield  {journal} {\bibinfo
  {journal} {J. Phys. Chem. Lett.}\ }\textbf {\bibinfo {volume} {13}},\
  \bibinfo {pages} {3890} (\bibinfo {year} {2022}{\natexlab{b}})}\BibitemShut
  {NoStop}%
\bibitem [{\citenamefont {Wang}\ and\ \citenamefont
  {Yelin}(2021)}]{wang2021roadmap}%
  \BibitemOpen
  \bibfield  {author} {\bibinfo {author} {\bibfnamefont {D.~S.}\ \bibnamefont
  {Wang}}\ and\ \bibinfo {author} {\bibfnamefont {S.~F.}\ \bibnamefont
  {Yelin}},\ }\bibfield  {title} {\bibinfo {title} {A roadmap toward the theory
  of vibrational polariton chemistry},\ }\href@noop {} {\bibfield  {journal}
  {\bibinfo  {journal} {ACS Photonics}\ }\textbf {\bibinfo {volume} {8}},\
  \bibinfo {pages} {2818} (\bibinfo {year} {2021})}\BibitemShut {NoStop}%
\bibitem [{\citenamefont {Li}\ \emph {et~al.}(2022{\natexlab{c}})\citenamefont
  {Li}, \citenamefont {Cui}, \citenamefont {Subotnik},\ and\ \citenamefont
  {Nitzan}}]{li2022review}%
  \BibitemOpen
  \bibfield  {author} {\bibinfo {author} {\bibfnamefont {T.~E.}\ \bibnamefont
  {Li}}, \bibinfo {author} {\bibfnamefont {B.}~\bibnamefont {Cui}}, \bibinfo
  {author} {\bibfnamefont {J.~E.}\ \bibnamefont {Subotnik}},\ and\ \bibinfo
  {author} {\bibfnamefont {A.}~\bibnamefont {Nitzan}},\ }\bibfield  {title}
  {\bibinfo {title} {Molecular polaritonics: Chemical dynamics under strong
  light–matter coupling},\ }\href@noop {} {\bibfield  {journal} {\bibinfo
  {journal} {Annu. Rev. Phys. Chem.}\ }\textbf {\bibinfo {volume} {73}},\
  \bibinfo {pages} {43} (\bibinfo {year} {2022}{\natexlab{c}})}\BibitemShut
  {NoStop}%
\bibitem [{\citenamefont {Sidler}\ \emph {et~al.}(2022)\citenamefont {Sidler},
  \citenamefont {Ruggenthaler}, \citenamefont {Sch{\"a}fer}, \citenamefont
  {Ronca},\ and\ \citenamefont {Rubio}}]{sidler2022perspective}%
  \BibitemOpen
  \bibfield  {author} {\bibinfo {author} {\bibfnamefont {D.}~\bibnamefont
  {Sidler}}, \bibinfo {author} {\bibfnamefont {M.}~\bibnamefont
  {Ruggenthaler}}, \bibinfo {author} {\bibfnamefont {C.}~\bibnamefont
  {Sch{\"a}fer}}, \bibinfo {author} {\bibfnamefont {E.}~\bibnamefont {Ronca}},\
  and\ \bibinfo {author} {\bibfnamefont {A.}~\bibnamefont {Rubio}},\ }\bibfield
   {title} {\bibinfo {title} {A perspective on \textit{ab initio} modeling of
  polaritonic chemistry: The role of non-equilibrium effects and quantum
  collectivity},\ }\href@noop {} {\bibfield  {journal} {\bibinfo  {journal} {J.
  Chem. Phys.}\ }\textbf {\bibinfo {volume} {156}},\ \bibinfo {pages} {230901}
  (\bibinfo {year} {2022})}\BibitemShut {NoStop}%
\bibitem [{\citenamefont {Mandal}\ \emph
  {et~al.}(2022{\natexlab{b}})\citenamefont {Mandal}, \citenamefont {Taylor},
  \citenamefont {Weight}, \citenamefont {Koessler}, \citenamefont {Li},\ and\
  \citenamefont {Huo}}]{mandal2022theoretical}%
  \BibitemOpen
  \bibfield  {author} {\bibinfo {author} {\bibfnamefont {A.}~\bibnamefont
  {Mandal}}, \bibinfo {author} {\bibfnamefont {M.}~\bibnamefont {Taylor}},
  \bibinfo {author} {\bibfnamefont {B.}~\bibnamefont {Weight}}, \bibinfo
  {author} {\bibfnamefont {E.}~\bibnamefont {Koessler}}, \bibinfo {author}
  {\bibfnamefont {X.}~\bibnamefont {Li}},\ and\ \bibinfo {author}
  {\bibfnamefont {P.}~\bibnamefont {Huo}},\ }\bibfield  {title} {\bibinfo
  {title} {Theoretical advances in polariton chemistry and molecular cavity
  quantum electrodynamics},\ }\href@noop {} {\bibfield  {journal} {\bibinfo
  {journal} {chemrxiv-2022-g9lr7}\ } (\bibinfo {year}
  {2022}{\natexlab{b}})}\BibitemShut {NoStop}%
\bibitem [{\citenamefont {Campos-Gonzalez-Angulo}\ \emph
  {et~al.}(2022)\citenamefont {Campos-Gonzalez-Angulo}, \citenamefont {Poh},
  \citenamefont {Du},\ and\ \citenamefont
  {Yuen-Zhou}}]{yuenzhou2022theoretical}%
  \BibitemOpen
  \bibfield  {author} {\bibinfo {author} {\bibfnamefont {J.~A.}\ \bibnamefont
  {Campos-Gonzalez-Angulo}}, \bibinfo {author} {\bibfnamefont {Y.~R.}\
  \bibnamefont {Poh}}, \bibinfo {author} {\bibfnamefont {M.}~\bibnamefont
  {Du}},\ and\ \bibinfo {author} {\bibfnamefont {J.}~\bibnamefont
  {Yuen-Zhou}},\ }\bibfield  {title} {\bibinfo {title} {Swinging between shine
  and shadow: Theoretical advances on thermally-activated vibropolaritonic
  chemistry (a perspective)},\ }\href@noop {} {\bibfield  {journal} {\bibinfo
  {journal} {arXiv:2212.04017}\ } (\bibinfo {year} {2022})}\BibitemShut
  {NoStop}%
\bibitem [{\citenamefont {Imperatore}\ \emph {et~al.}(2021)\citenamefont
  {Imperatore}, \citenamefont {Asbury},\ and\ \citenamefont
  {Giebink}}]{imperatore2021reproducibility}%
  \BibitemOpen
  \bibfield  {author} {\bibinfo {author} {\bibfnamefont {M.~V.}\ \bibnamefont
  {Imperatore}}, \bibinfo {author} {\bibfnamefont {J.~B.}\ \bibnamefont
  {Asbury}},\ and\ \bibinfo {author} {\bibfnamefont {N.~C.}\ \bibnamefont
  {Giebink}},\ }\bibfield  {title} {\bibinfo {title} {Reproducibility of
  cavity-enhanced chemical reaction rates in the vibrational strong coupling
  regime},\ }\href@noop {} {\bibfield  {journal} {\bibinfo  {journal} {J. Chem.
  Phys.}\ }\textbf {\bibinfo {volume} {154}},\ \bibinfo {pages} {191103}
  (\bibinfo {year} {2021})}\BibitemShut {NoStop}%
\bibitem [{\citenamefont {Miller}(1978)}]{miller1978classical}%
  \BibitemOpen
  \bibfield  {author} {\bibinfo {author} {\bibfnamefont {W.~H.}\ \bibnamefont
  {Miller}},\ }\bibfield  {title} {\bibinfo {title} {A classical/semiclassical
  theory for the interaction of infrared radiation with molecular systems},\
  }\href@noop {} {\bibfield  {journal} {\bibinfo  {journal} {J. Chem. Phys.}\
  }\textbf {\bibinfo {volume} {69}},\ \bibinfo {pages} {2188} (\bibinfo {year}
  {1978})}\BibitemShut {NoStop}%
\bibitem [{\citenamefont {Brown}\ and\ \citenamefont
  {Wyatt}(1986{\natexlab{a}})}]{brown1986quantum}%
  \BibitemOpen
  \bibfield  {author} {\bibinfo {author} {\bibfnamefont {R.~C.}\ \bibnamefont
  {Brown}}\ and\ \bibinfo {author} {\bibfnamefont {R.~E.}\ \bibnamefont
  {Wyatt}},\ }\bibfield  {title} {\bibinfo {title} {Quantum mechanical
  manifestation of cantori: Wave-packet localization in stochastic regions},\
  }\href@noop {} {\bibfield  {journal} {\bibinfo  {journal} {Phys. Rev. Lett.}\
  }\textbf {\bibinfo {volume} {57}},\ \bibinfo {pages} {1} (\bibinfo {year}
  {1986}{\natexlab{a}})}\BibitemShut {NoStop}%
\bibitem [{\citenamefont {Shirts}\ and\ \citenamefont
  {Davis}(1984)}]{shirts1984cm}%
  \BibitemOpen
  \bibfield  {author} {\bibinfo {author} {\bibfnamefont {R.~B.}\ \bibnamefont
  {Shirts}}\ and\ \bibinfo {author} {\bibfnamefont {T.~F.}\ \bibnamefont
  {Davis}},\ }\bibfield  {title} {\bibinfo {title} {Classical resonance
  analysis in conservative models of infrared absorption},\ }\href@noop {}
  {\bibfield  {journal} {\bibinfo  {journal} {J. Phys. Chem.}\ }\textbf
  {\bibinfo {volume} {88}},\ \bibinfo {pages} {4665} (\bibinfo {year}
  {1984})}\BibitemShut {NoStop}%
\bibitem [{\citenamefont {Davis}\ and\ \citenamefont
  {Wyatt}(1982)}]{davis1982surface}%
  \BibitemOpen
  \bibfield  {author} {\bibinfo {author} {\bibfnamefont {M.~J.}\ \bibnamefont
  {Davis}}\ and\ \bibinfo {author} {\bibfnamefont {R.~E.}\ \bibnamefont
  {Wyatt}},\ }\bibfield  {title} {\bibinfo {title} {Surface-of-section analysis
  in the classical theory of multiphoton absorption},\ }\href@noop {}
  {\bibfield  {journal} {\bibinfo  {journal} {Chem. Phys. Lett.}\ }\textbf
  {\bibinfo {volume} {86}},\ \bibinfo {pages} {235} (\bibinfo {year}
  {1982})}\BibitemShut {NoStop}%
\bibitem [{\citenamefont {Fischer}\ and\ \citenamefont
  {Saalfrank}(2021)}]{fischer2021ground}%
  \BibitemOpen
  \bibfield  {author} {\bibinfo {author} {\bibfnamefont {E.~W.}\ \bibnamefont
  {Fischer}}\ and\ \bibinfo {author} {\bibfnamefont {P.}~\bibnamefont
  {Saalfrank}},\ }\bibfield  {title} {\bibinfo {title} {Ground state properties
  and infrared spectra of anharmonic vibrational polaritons of small molecules
  in cavities},\ }\href@noop {} {\bibfield  {journal} {\bibinfo  {journal} {J.
  Chem. Phys.}\ }\textbf {\bibinfo {volume} {154}},\ \bibinfo {pages} {104311}
  (\bibinfo {year} {2021})}\BibitemShut {NoStop}%
\bibitem [{\citenamefont {Brown}\ and\ \citenamefont
  {Wyatt}(1986{\natexlab{b}})}]{brown1986barriers}%
  \BibitemOpen
  \bibfield  {author} {\bibinfo {author} {\bibfnamefont {R.~C.}\ \bibnamefont
  {Brown}}\ and\ \bibinfo {author} {\bibfnamefont {R.~E.}\ \bibnamefont
  {Wyatt}},\ }\bibfield  {title} {\bibinfo {title} {Barriers to chaotic
  classical motion and quantum mechanical localization in multiphoton
  dissociation},\ }\href@noop {} {\bibfield  {journal} {\bibinfo  {journal} {J.
  Phys. Chem.}\ }\textbf {\bibinfo {volume} {90}},\ \bibinfo {pages} {3590}
  (\bibinfo {year} {1986}{\natexlab{b}})}\BibitemShut {NoStop}%
\bibitem [{\citenamefont {Sch\"{a}fer}\ \emph {et~al.}(2020)\citenamefont
  {Sch\"{a}fer}, \citenamefont {Ruggenthaler}, \citenamefont {Rokaj},\ and\
  \citenamefont {Rubio}}]{schafer2020relevance}%
  \BibitemOpen
  \bibfield  {author} {\bibinfo {author} {\bibfnamefont {C.}~\bibnamefont
  {Sch\"{a}fer}}, \bibinfo {author} {\bibfnamefont {M.}~\bibnamefont
  {Ruggenthaler}}, \bibinfo {author} {\bibfnamefont {V.}~\bibnamefont
  {Rokaj}},\ and\ \bibinfo {author} {\bibfnamefont {A.}~\bibnamefont {Rubio}},\
  }\bibfield  {title} {\bibinfo {title} {Relevance of the quadratic diamagnetic
  and self-polarization terms in cavity quantum electrodynamics},\ }\href@noop
  {} {\bibfield  {journal} {\bibinfo  {journal} {ACS photonics}\ }\textbf
  {\bibinfo {volume} {7}},\ \bibinfo {pages} {975} (\bibinfo {year}
  {2020})}\BibitemShut {NoStop}%
\bibitem [{\citenamefont {Stine}\ and\ \citenamefont
  {Noid}(1979)}]{stine1979classical}%
  \BibitemOpen
  \bibfield  {author} {\bibinfo {author} {\bibfnamefont {J.}~\bibnamefont
  {Stine}}\ and\ \bibinfo {author} {\bibfnamefont {D.}~\bibnamefont {Noid}},\
  }\bibfield  {title} {\bibinfo {title} {Classical treatment of the
  dissociation of hydrogen fluoride with one and two infrared lasers},\
  }\href@noop {} {\bibfield  {journal} {\bibinfo  {journal} {Opt. Commun.}\
  }\textbf {\bibinfo {volume} {31}},\ \bibinfo {pages} {161} (\bibinfo {year}
  {1979})}\BibitemShut {NoStop}%
\bibitem [{\citenamefont {de~Lima}\ \emph {et~al.}(2013)\citenamefont
  {de~Lima}, \citenamefont {Ramos},\ and\ \citenamefont
  {de~Carvalho}}]{lima2013cm}%
  \BibitemOpen
  \bibfield  {author} {\bibinfo {author} {\bibfnamefont {E.~F.}\ \bibnamefont
  {de~Lima}}, \bibinfo {author} {\bibfnamefont {T.~N.}\ \bibnamefont {Ramos}},\
  and\ \bibinfo {author} {\bibfnamefont {R.~E.}\ \bibnamefont {de~Carvalho}},\
  }\bibfield  {title} {\bibinfo {title} {Role of the range of the dipole
  function in the classical dynamics of molecular dissociation},\ }\href@noop
  {} {\bibfield  {journal} {\bibinfo  {journal} {Phys. Rev. E}\ }\textbf
  {\bibinfo {volume} {87}},\ \bibinfo {pages} {014901} (\bibinfo {year}
  {2013})}\BibitemShut {NoStop}%
\bibitem [{\citenamefont {de~Lima}\ \emph {et~al.}(2014)\citenamefont
  {de~Lima}, \citenamefont {Rosado}, \citenamefont {Castelano},\ and\
  \citenamefont {de~Carvalho}}]{lima2014qm}%
  \BibitemOpen
  \bibfield  {author} {\bibinfo {author} {\bibfnamefont {E.~F.}\ \bibnamefont
  {de~Lima}}, \bibinfo {author} {\bibfnamefont {E.}~\bibnamefont {Rosado}},
  \bibinfo {author} {\bibfnamefont {L.}~\bibnamefont {Castelano}},\ and\
  \bibinfo {author} {\bibfnamefont {R.~E.}\ \bibnamefont {de~Carvalho}},\
  }\bibfield  {title} {\bibinfo {title} {Quantum–classical correspondence and
  the role of the dipole function in molecular dissociation},\ }\href@noop {}
  {\bibfield  {journal} {\bibinfo  {journal} {Phys. Lett. A}\ }\textbf
  {\bibinfo {volume} {378}},\ \bibinfo {pages} {2657} (\bibinfo {year}
  {2014})}\BibitemShut {NoStop}%
\bibitem [{\citenamefont {Triana}\ \emph {et~al.}(2020)\citenamefont {Triana},
  \citenamefont {Hern{\'a}ndez},\ and\ \citenamefont
  {Herrera}}]{triana2020shape}%
  \BibitemOpen
  \bibfield  {author} {\bibinfo {author} {\bibfnamefont {J.~F.}\ \bibnamefont
  {Triana}}, \bibinfo {author} {\bibfnamefont {F.~J.}\ \bibnamefont
  {Hern{\'a}ndez}},\ and\ \bibinfo {author} {\bibfnamefont {F.}~\bibnamefont
  {Herrera}},\ }\bibfield  {title} {\bibinfo {title} {The shape of the electric
  dipole function determines the sub-picosecond dynamics of anharmonic
  vibrational polaritons},\ }\href@noop {} {\bibfield  {journal} {\bibinfo
  {journal} {J. Chem. Phys.}\ }\textbf {\bibinfo {volume} {152}},\ \bibinfo
  {pages} {234111} (\bibinfo {year} {2020})}\BibitemShut {NoStop}%
\bibitem [{\citenamefont {Chirikov}(1979)}]{chirikov1979universal}%
  \BibitemOpen
  \bibfield  {author} {\bibinfo {author} {\bibfnamefont {B.~V.}\ \bibnamefont
  {Chirikov}},\ }\bibfield  {title} {\bibinfo {title} {A universal instability
  of many-dimensional oscillator systems},\ }\href@noop {} {\bibfield
  {journal} {\bibinfo  {journal} {Phys. Rep.}\ }\textbf {\bibinfo {volume}
  {52}},\ \bibinfo {pages} {263} (\bibinfo {year} {1979})}\BibitemShut
  {NoStop}%
\bibitem [{\citenamefont {Sethi}\ and\ \citenamefont
  {Keshavamurthy}(2009)}]{sethi2009local}%
  \BibitemOpen
  \bibfield  {author} {\bibinfo {author} {\bibfnamefont {A.}~\bibnamefont
  {Sethi}}\ and\ \bibinfo {author} {\bibfnamefont {S.}~\bibnamefont
  {Keshavamurthy}},\ }\bibfield  {title} {\bibinfo {title} {Local phase space
  control and interplay of classical and quantum effects in dissociation of a
  driven {M}orse oscillator},\ }\href@noop {} {\bibfield  {journal} {\bibinfo
  {journal} {Phys. Rev. A}\ }\textbf {\bibinfo {volume} {79}},\ \bibinfo
  {pages} {033416} (\bibinfo {year} {2009})}\BibitemShut {NoStop}%
\bibitem [{\citenamefont {Shirts}(1987)}]{shirts1987use}%
  \BibitemOpen
  \bibfield  {author} {\bibinfo {author} {\bibfnamefont {R.~B.}\ \bibnamefont
  {Shirts}},\ }\bibfield  {title} {\bibinfo {title} {Use of classical {F}ourier
  amplitudes as quantum matrix elements: a comparison of {M}orse oscillator
  {F}ourier coefficients with quantum matrix elements},\ }\href@noop {}
  {\bibfield  {journal} {\bibinfo  {journal} {J. Phys. Chem.}\ }\textbf
  {\bibinfo {volume} {91}},\ \bibinfo {pages} {2258} (\bibinfo {year}
  {1987})}\BibitemShut {NoStop}%
\bibitem [{\citenamefont {Triana}\ and\ \citenamefont
  {Herrera}(2020)}]{triana2020self}%
  \BibitemOpen
  \bibfield  {author} {\bibinfo {author} {\bibfnamefont {J.}~\bibnamefont
  {Triana}}\ and\ \bibinfo {author} {\bibfnamefont {F.}~\bibnamefont
  {Herrera}},\ }\bibfield  {title} {\bibinfo {title} {Self-dissociation of
  polar molecules in a confined infrared vacuum},\ }\href@noop {} {\bibfield
  {journal} {\bibinfo  {journal} {chemrxiv.12702419.v1}\ } (\bibinfo {year}
  {2020})}\BibitemShut {NoStop}%
\bibitem [{\citenamefont {Wigner}(1938)}]{wigner1938transition}%
  \BibitemOpen
  \bibfield  {author} {\bibinfo {author} {\bibfnamefont {E.}~\bibnamefont
  {Wigner}},\ }\bibfield  {title} {\bibinfo {title} {The transition state
  method},\ }\href@noop {} {\bibfield  {journal} {\bibinfo  {journal} {Trans.
  Faraday Soc.}\ }\textbf {\bibinfo {volume} {34}},\ \bibinfo {pages} {29}
  (\bibinfo {year} {1938})}\BibitemShut {NoStop}%
\bibitem [{\citenamefont {Waalkens}\ \emph {et~al.}(2007)\citenamefont
  {Waalkens}, \citenamefont {Schubert},\ and\ \citenamefont
  {Wiggins}}]{waalkens2007wigner}%
  \BibitemOpen
  \bibfield  {author} {\bibinfo {author} {\bibfnamefont {H.}~\bibnamefont
  {Waalkens}}, \bibinfo {author} {\bibfnamefont {R.}~\bibnamefont {Schubert}},\
  and\ \bibinfo {author} {\bibfnamefont {S.}~\bibnamefont {Wiggins}},\
  }\bibfield  {title} {\bibinfo {title} {Wigner's dynamical transition state
  theory in phase space: classical and quantum},\ }\href@noop {} {\bibfield
  {journal} {\bibinfo  {journal} {Nonlinearity}\ }\textbf {\bibinfo {volume}
  {21}},\ \bibinfo {pages} {R1} (\bibinfo {year} {2007})}\BibitemShut {NoStop}%
\bibitem [{\citenamefont {Jaff{\'e}}\ \emph {et~al.}(2005)\citenamefont
  {Jaff{\'e}}, \citenamefont {Kawai}, \citenamefont {Palaci{\'a}n},
  \citenamefont {Yanguas},\ and\ \citenamefont {Uzer}}]{uzertstadvchemphys}%
  \BibitemOpen
  \bibfield  {author} {\bibinfo {author} {\bibfnamefont {C.}~\bibnamefont
  {Jaff{\'e}}}, \bibinfo {author} {\bibfnamefont {S.}~\bibnamefont {Kawai}},
  \bibinfo {author} {\bibfnamefont {J.}~\bibnamefont {Palaci{\'a}n}}, \bibinfo
  {author} {\bibfnamefont {P.}~\bibnamefont {Yanguas}},\ and\ \bibinfo {author}
  {\bibfnamefont {T.}~\bibnamefont {Uzer}},\ }\bibfield  {title} {\bibinfo
  {title} {A new look at the transition state: {W}igner's dynamical perspective
  revisited},\ }\href@noop {} {\bibfield  {journal} {\bibinfo  {journal} {Adv.
  Chem. Phys.}\ }\textbf {\bibinfo {volume} {130A}},\ \bibinfo {pages} {171}
  (\bibinfo {year} {2005})}\BibitemShut {NoStop}%
\bibitem [{\citenamefont {Zhao}\ \emph {et~al.}(2005)\citenamefont {Zhao},
  \citenamefont {Gong},\ and\ \citenamefont {Rice}}]{zhao2005classical}%
  \BibitemOpen
  \bibfield  {author} {\bibinfo {author} {\bibfnamefont {M.}~\bibnamefont
  {Zhao}}, \bibinfo {author} {\bibfnamefont {J.}~\bibnamefont {Gong}},\ and\
  \bibinfo {author} {\bibfnamefont {S.~A.}\ \bibnamefont {Rice}},\ }\bibfield
  {title} {\bibinfo {title} {Classical, semiclassical, and quantum mechanical
  unimolecular reaction rate theory},\ }\href@noop {} {\bibfield  {journal}
  {\bibinfo  {journal} {Adv. Chem. Phys.}\ }\textbf {\bibinfo {volume}
  {130A}},\ \bibinfo {pages} {1} (\bibinfo {year} {2005})}\BibitemShut
  {NoStop}%
\bibitem [{\citenamefont {Ribeiro}(2022)}]{ribeiro2022multimode}%
  \BibitemOpen
  \bibfield  {author} {\bibinfo {author} {\bibfnamefont {R.~F.}\ \bibnamefont
  {Ribeiro}},\ }\bibfield  {title} {\bibinfo {title} {Multimode polariton
  effects on molecular energy transport and spectral fluctuations},\
  }\href@noop {} {\bibfield  {journal} {\bibinfo  {journal} {Commun. Chem.}\
  }\textbf {\bibinfo {volume} {5}},\ \bibinfo {pages} {1} (\bibinfo {year}
  {2022})}\BibitemShut {NoStop}%
\bibitem [{\citenamefont {Vurgaftman}\ \emph {et~al.}(2022)\citenamefont
  {Vurgaftman}, \citenamefont {Simpkins}, \citenamefont {Dunkelberger},\ and\
  \citenamefont {Owrutsky}}]{vurgaftman2022comparative}%
  \BibitemOpen
  \bibfield  {author} {\bibinfo {author} {\bibfnamefont {I.}~\bibnamefont
  {Vurgaftman}}, \bibinfo {author} {\bibfnamefont {B.~S.}\ \bibnamefont
  {Simpkins}}, \bibinfo {author} {\bibfnamefont {A.~D.}\ \bibnamefont
  {Dunkelberger}},\ and\ \bibinfo {author} {\bibfnamefont {J.~C.}\ \bibnamefont
  {Owrutsky}},\ }\bibfield  {title} {\bibinfo {title} {Comparative analysis of
  polaritons in bulk, dielectric slabs, and planar cavities with implications
  for cavity-modified reactivity},\ }\href@noop {} {\bibfield  {journal}
  {\bibinfo  {journal} {J. Chem. Phys.}\ }\textbf {\bibinfo {volume} {156}},\
  \bibinfo {pages} {034110} (\bibinfo {year} {2022})}\BibitemShut {NoStop}%
\bibitem [{\citenamefont {Zhou}\ \emph {et~al.}(2022)\citenamefont {Zhou},
  \citenamefont {Chen}, \citenamefont {Subotnik},\ and\ \citenamefont
  {Nitzan}}]{zhou2022interplay}%
  \BibitemOpen
  \bibfield  {author} {\bibinfo {author} {\bibfnamefont {Z.}~\bibnamefont
  {Zhou}}, \bibinfo {author} {\bibfnamefont {H.-T.}\ \bibnamefont {Chen}},
  \bibinfo {author} {\bibfnamefont {J.~E.}\ \bibnamefont {Subotnik}},\ and\
  \bibinfo {author} {\bibfnamefont {A.}~\bibnamefont {Nitzan}},\ }\bibfield
  {title} {\bibinfo {title} {The interplay between disorder, local relaxation
  and collective behaviors for an ensemble of emitters outside vs inside
  cavity},\ }\href@noop {} {\bibfield  {journal} {\bibinfo  {journal}
  {arXiv:2211.16325}\ } (\bibinfo {year} {2022})}\BibitemShut {NoStop}%
\bibitem [{\citenamefont {Li}\ \emph {et~al.}(2022{\natexlab{d}})\citenamefont
  {Li}, \citenamefont {Nitzan},\ and\ \citenamefont {Subotnik}}]{li2022energy}%
  \BibitemOpen
  \bibfield  {author} {\bibinfo {author} {\bibfnamefont {T.~E.}\ \bibnamefont
  {Li}}, \bibinfo {author} {\bibfnamefont {A.}~\bibnamefont {Nitzan}},\ and\
  \bibinfo {author} {\bibfnamefont {J.~E.}\ \bibnamefont {Subotnik}},\
  }\bibfield  {title} {\bibinfo {title} {{Energy-efficient pathway for
  selectively exciting solute molecules to high vibrational states via solvent
  vibration-polariton pumping}},\ }\href@noop {} {\bibfield  {journal}
  {\bibinfo  {journal} {Nat. Commun.}\ }\textbf {\bibinfo {volume} {13}},\
  \bibinfo {pages} {1} (\bibinfo {year} {2022}{\natexlab{d}})}\BibitemShut
  {NoStop}%
\bibitem [{\citenamefont {Fischer}\ \emph {et~al.}(2022)\citenamefont
  {Fischer}, \citenamefont {Anders},\ and\ \citenamefont
  {Saalfrank}}]{fischer2022cavity}%
  \BibitemOpen
  \bibfield  {author} {\bibinfo {author} {\bibfnamefont {E.~W.}\ \bibnamefont
  {Fischer}}, \bibinfo {author} {\bibfnamefont {J.}~\bibnamefont {Anders}},\
  and\ \bibinfo {author} {\bibfnamefont {P.}~\bibnamefont {Saalfrank}},\
  }\bibfield  {title} {\bibinfo {title} {{Cavity-altered thermal isomerization
  rates and dynamical resonant localization in vibro-polaritonic chemistry}},\
  }\href@noop {} {\bibfield  {journal} {\bibinfo  {journal} {J. Chem. Phys.}\
  }\textbf {\bibinfo {volume} {156}},\ \bibinfo {pages} {154305} (\bibinfo
  {year} {2022})}\BibitemShut {NoStop}%
\bibitem [{\citenamefont {Sethi}\ and\ \citenamefont
  {Keshavamurthy}(2012)}]{ASKSmolphys}%
  \BibitemOpen
  \bibfield  {author} {\bibinfo {author} {\bibfnamefont {A.}~\bibnamefont
  {Sethi}}\ and\ \bibinfo {author} {\bibfnamefont {S.}~\bibnamefont
  {Keshavamurthy}},\ }\bibfield  {title} {\bibinfo {title} {Driven coupled
  {M}orse oscillators: visualizing the phase space and characterizing the
  transport},\ }\href {https://doi.org/10.1080/00268976.2012.667166} {\bibfield
   {journal} {\bibinfo  {journal} {Mol. Phys.}\ }\textbf {\bibinfo {volume}
  {110}},\ \bibinfo {pages} {717} (\bibinfo {year} {2012})}\BibitemShut
  {NoStop}%
\bibitem [{\citenamefont {Lopez-Pina}\ \emph {et~al.}(2016)\citenamefont
  {Lopez-Pina}, \citenamefont {Losada}, \citenamefont {Benito},\ and\
  \citenamefont {Borondo}}]{BorondoHCN}%
  \BibitemOpen
  \bibfield  {author} {\bibinfo {author} {\bibfnamefont {A.}~\bibnamefont
  {Lopez-Pina}}, \bibinfo {author} {\bibfnamefont {J.~C.}\ \bibnamefont
  {Losada}}, \bibinfo {author} {\bibfnamefont {R.~M.}\ \bibnamefont {Benito}},\
  and\ \bibinfo {author} {\bibfnamefont {F.}~\bibnamefont {Borondo}},\
  }\bibfield  {title} {\bibinfo {title} {Frequency analysis of the laser driven
  nonlinear dynamics of {HCN}},\ }\href {https://doi.org/10.1063/1.4972260}
  {\bibfield  {journal} {\bibinfo  {journal} {J. Chem. Phys.}\ }\textbf
  {\bibinfo {volume} {145}},\ \bibinfo {pages} {244309} (\bibinfo {year}
  {2016})}\BibitemShut {NoStop}%
\bibitem [{\citenamefont {Karmakar}\ \emph {et~al.}(2020)\citenamefont
  {Karmakar}, \citenamefont {Yadav},\ and\ \citenamefont
  {Keshavamurthy}}]{SKPKYKS2020}%
  \BibitemOpen
  \bibfield  {author} {\bibinfo {author} {\bibfnamefont {S.}~\bibnamefont
  {Karmakar}}, \bibinfo {author} {\bibfnamefont {P.~K.}\ \bibnamefont
  {Yadav}},\ and\ \bibinfo {author} {\bibfnamefont {S.}~\bibnamefont
  {Keshavamurthy}},\ }\bibfield  {title} {\bibinfo {title} {Stable chaos and
  delayed onset of statisticality in unimolecular dissociation reactions},\
  }\href@noop {} {\bibfield  {journal} {\bibinfo  {journal} {Commun. Chem.}\
  }\textbf {\bibinfo {volume} {3}},\ \bibinfo {pages} {4} (\bibinfo {year}
  {2020})}\BibitemShut {NoStop}%
\bibitem [{\citenamefont {Semparithi}\ and\ \citenamefont
  {Keshavamurthy}(2006)}]{Semparithi2006}%
  \BibitemOpen
  \bibfield  {author} {\bibinfo {author} {\bibfnamefont {A.}~\bibnamefont
  {Semparithi}}\ and\ \bibinfo {author} {\bibfnamefont {S.}~\bibnamefont
  {Keshavamurthy}},\ }\bibfield  {title} {\bibinfo {title} {Intramolecular
  vibrational energy redistribution as state space diffusion: Classical-quantum
  correspondence},\ }\href {https://doi.org/10.1063/1.2358138} {\bibfield
  {journal} {\bibinfo  {journal} {J. Chem. Phys.}\ }\textbf {\bibinfo {volume}
  {125}},\ \bibinfo {pages} {141101} (\bibinfo {year} {2006})}\BibitemShut
  {NoStop}%
\bibitem [{\citenamefont {Manikandan}\ and\ \citenamefont
  {Keshavamurthy}(2014)}]{ManikandanKS}%
  \BibitemOpen
  \bibfield  {author} {\bibinfo {author} {\bibfnamefont {P.}~\bibnamefont
  {Manikandan}}\ and\ \bibinfo {author} {\bibfnamefont {S.}~\bibnamefont
  {Keshavamurthy}},\ }\bibfield  {title} {\bibinfo {title} {Dynamical traps
  lead to the slowing down of intramolecular vibrational energy flow},\
  }\href@noop {} {\bibfield  {journal} {\bibinfo  {journal} {Proc. Natl. Acad.
  Sci. U.S.A.}\ }\textbf {\bibinfo {volume} {111}},\ \bibinfo {pages} {14354}
  (\bibinfo {year} {2014})}\BibitemShut {NoStop}%
\bibitem [{\citenamefont {Feit}\ and\ \citenamefont
  {Fleck~Jr}(1983)}]{feit1983solution}%
  \BibitemOpen
  \bibfield  {author} {\bibinfo {author} {\bibfnamefont {M.}~\bibnamefont
  {Feit}}\ and\ \bibinfo {author} {\bibfnamefont {J.}~\bibnamefont
  {Fleck~Jr}},\ }\bibfield  {title} {\bibinfo {title} {{Solution of the
  Schr{\"o}dinger equation by a spectral method II: Vibrational energy levels
  of triatomic molecules}},\ }\href@noop {} {\bibfield  {journal} {\bibinfo
  {journal} {J. Chem. Phys.}\ }\textbf {\bibinfo {volume} {78}},\ \bibinfo
  {pages} {301} (\bibinfo {year} {1983})}\BibitemShut {NoStop}%
\bibitem [{\citenamefont {Leforestier}\ and\ \citenamefont
  {Wyatt}(1983)}]{leforestier1983optical}%
  \BibitemOpen
  \bibfield  {author} {\bibinfo {author} {\bibfnamefont {C.}~\bibnamefont
  {Leforestier}}\ and\ \bibinfo {author} {\bibfnamefont {R.~E.}\ \bibnamefont
  {Wyatt}},\ }\bibfield  {title} {\bibinfo {title} {Optical potential for laser
  induced dissociation},\ }\href@noop {} {\bibfield  {journal} {\bibinfo
  {journal} {J. Chem. Phys.}\ }\textbf {\bibinfo {volume} {78}},\ \bibinfo
  {pages} {2334} (\bibinfo {year} {1983})}\BibitemShut {NoStop}%
\end{thebibliography}%

\end{document}